\documentclass[fleqn,usenatbib]{mnras}

\usepackage{newtxtext,newtxmath}
\usepackage[T1]{fontenc}

\DeclareRobustCommand{\VAN}[3]{#2}
\let\VANthebibliography\thebibliography
\def\thebibliography{\DeclareRobustCommand{\VAN}[3]{##3}\VANthebibliography}

\usepackage{graphicx}	% Including figure files
\usepackage{amsmath}	% Advanced maths commands
\usepackage{algorithmic}
\usepackage{booktabs}
\usepackage{natbib}
\usepackage{caption}
\usepackage{subcaption}
\graphicspath{ {figures/} }

\defcitealias{Banik2019}{Paper I}

\title[SMBH formation and evolution to the local universe]{The formation of supermassive black holes from Population III.1 seeds. II. Evolution to the local universe}

\author[Singh et al.]{
Jasbir Singh,$^{1,2,3,4}$\thanks{E-mail: jasbir.singh@inaf.it}
Pierluigi Monaco,$^{1,2,3,5}$
Jonathan C. Tan$^{4,6}$
\\
$^{1}$Astronomy Unit, Department of Physics, University of Trieste, via Tiepolo 11, I-34131 Trieste, Italy\\
$^{2}$INAF- Astronomical Observatory of Trieste, via Tiepolo 11, 34143 Trieste, Italy\\
$^{3}$IFPU – Institute for Fundamental Physics of the Universe, Via Beirut 2, I-34014 Trieste, Italy\\
$^{4}$Department of Space, Earth \& Environment, Chalmers University of Technology, Gothenburg, Sweden\\
$^{5}$INFN, Sezione di Trieste, 34149 Trieste, Italy \\
$^{6}$Dept. of Astronomy, University of Virginia, Charlottesville, VA 22904, USA
}

\date{Accepted 27 July 2023}

\pubyear{2023}

\begin{document}
\label{firstpage}
\pagerange{\pageref{firstpage}--\pageref{lastpage}}
\maketitle

\begin{abstract}
We present predictions for cosmic evolution of populations of supermassive black holes (SMBHs) forming from Population III.1 seeds, i.e., early, metal-free dark matter minihalos forming far from other sources, parameterized by isolation distance, $d_{\rm{iso}}$. Extending previous work that explored this scenario to $z=10$, we follow evolution of a $(60\:{\rm{Mpc}})^3$ volume to $z=0$. We focus on evolution of SMBH comoving number densities, halo occupation fractions, angular clustering and 3D clustering, exploring a range of $d_{\rm{iso}}$ constrained by observed local number densities of SMBHs. We also compute synthetic projected observational fields, in particular a case comparable to the Hubble Ultra Deep Field. We compare Pop III.1 seeding to a simple halo mass threshold model, commonly adopted in cosmological simulations of galaxy formation. Major predictions of the Pop III.1 model include that all SMBHs form by $z\sim25$, after which their comoving number densities are near-constant, with low merger rates. Occupation fractions evolve to concentrate SMBHs in the most massive halos by $z=0$, but with rare cases of SMBHs in halos down to $\sim10^8\:M_\odot$. The $d_{\rm{iso}}$ scale at epoch of formation, e.g., $100\:$kpc-proper at $z\sim30$, i.e., $\sim3\:$Mpc-comoving, is imprinted in the SMBH two-point angular correlation function, remaining discernible as a low-amplitude feature to $z\sim1$. The SMBH 3D two-point correlation function at $z=0$ also shows lower amplitude compared to equivalently massive halos. We discuss prospects for testing these predictions with observational surveys of SMBH populations.
\end{abstract}


%%%%%%%%%%%%%%%%%%%%%%%%%%%%%%%%%%%%%%%%%%%%%%%%%%

%%%%%%%%%%%%%%%%% BODY OF PAPER %%%%%%%%%%%%%%%%%%

\section{Introduction}
\label{sec:introduction}

The origin of supermassive black holes (SMBHs) is one of the most outstanding open questions of contemporary astrophysics. These SMBHs have masses $\gtrsim10^5 M_\odot$ and are found at the center of most massive galaxies \citep[e.g.,][]{Graham16, Volonteri21, Lusso22}. 
Discoveries of high redshift quasars, such as J1007+2115 at $z=7.515$ \citep{Yang20} and J0313-1806 at $z=7.642$ \citep[][]{Wang21}, which are estimated to host SMBHs with masses $\gtrsim 10^9\:M_\odot$, place stringent constraints on SMBH formation and growth scenarios. In particular, the existence of these quasars imply that at least some SMBHs could form and grow efficiently to very high masses by the time the universe was only $\sim700$ million years old. Even assuming very early formation at $z\sim30$, for Eddington-limited accretion the SMBH seed mass would need to be $\gtrsim 10^4\:M_\odot$ and a later formation epoch would imply even higher seed masses. While scenarios of super-Eddington accretion have been proposed \citep[e.g.,][]{Kohri22}, numerical simulations indicate that typical gas supply rates to early-formed SMBHs are impacted by star formation feedback and will be far below the level needed to sustain Eddington-limited accretion rates \citep[e.g.,][]{Oshea05,Jeon23}. These considerations motivate the need for models of black hole formation at the supermassive, $\gtrsim10^5\:M_\odot$ scale.

There are a variety of proposed ideas for the physical mechanism of SMBH formation \citep[e.g.,][]{Rees78}. One suggested process is ``direct collapse'', which involves a massive primordial composition gas cloud contained in a relatively massive, atomically-cooled halo of $\sim 10^8 M_\odot$. The cloud collapses into a single, supermassive star of $10^4-10^6 M_\odot$ that then forms a SMBH \citep[e.g.,][]{Bromm03, Begelman06, Lodato06, Shang10, Montero12, Maio19, Bhowmick22a}. Although the number density of black holes emerging from direct collapse would be enough to explain the currently known population of high redshift quasars, the conditions required for this scenario are not thought to be common enough to explain the total observed population of SMBHs at $z=0$ \citep{Chon16,Wise2019}. Furthermore, recent simulations have shown that the supermassive stars forming via this mechanism might not be as massive as initially predicted, but only reaching $\lesssim 10^4 M_\odot$, due to the turbulent environment present in the initial stages of galaxy formation, which disrupts the accretion flow \citep{Regan20}. 

Another mechanism to form intermediate, or even supermassive black holes is through runaway stellar mergers in young and dense clusters to create stars with masses of the order $\sim 200-10^3 M_\odot$ \citep[e.g.,][]{Zwart04}. This mass can be reached through repeated collisions if the massive stars can reach the cluster core to increase the collision rate drastically \citep{Ebisuzaki03} before they explode as supernovae. Gas accretion driven compression of a dense cluster of stellar mass black holes to form a SMBH has also been proposed \citep{Kroupa20}. However, in general predicting whether the conditions needed for such dense clusters arise in galaxies and at what rate is very challenging given the the need to resolve the formation and evolution of individual stars, so predictions for the cosmological population of such systems are highly uncertain \citep[see, e.g.,][]{Boekholt18,Chon20,Tagawa20}.

Another class of SMBH seeding model considers the very first, so-called Population (Pop) III stars as potential progenitors. However, conventional models of Pop III star formation predict stellar masses that are ``only'' $\sim 100\:M_\odot$ \citep[e.g.,][]{Abel02,Bromm02,Tan04,MT08,Hosokawa11,Susa14}, which would only have the ability to produce stellar-mass black holes, i.e., relatively low-mass seeds. However, as discussed in more detail below, Pop III SMBH seeding models have been revived by \citet{Banik2019} when allowing for the potential effects of dark matter self-annihilation on the mass scale of formed stars \citep[][]{SFG08,NTO09,F10,RD15}.

More exotic models involving modification of the standard cold dark matter paradigm have also been proposed. For example, if dark matter undergoes self-interaction, then this could provide a mechanism for SMBH seeding via collapse of the halos themselves \citep[e.g.,][]{Feng21}. An even more extreme scenario is one in which SMBHs are primordial black holes, although this appears to be disfavoured by the clustering analysis of \citet[][]{Shinohara23}.

Given the uncertainty of SMBH formation models and the difficulty of resolving the small-scale physics, cosmological simulations have typically made very simplified assumptions for the SMBH seeding process based on the properties of the parent halo or galaxy. One of the simplest and most widely used models is the halo mass threshold (HMT) seeding scheme based on the methods developed by \citet{Sijacki07} and \citet{DiMatteo08}, in which a seed black hole is assumed to form in a halo crossing a certain mass threshold. The Illustris project \citep{Vogelsberger14} used this mechanism to add SMBHs of mass $1.4\times10^5 M_\odot$ in each halo which crosses a mass threshold of $m_{\rm th}=7.1\times10^{10} M_\odot$. A similar approach was used in the Evolution and Assembly of GaLaxies and their Environments (EAGLE) simulation \citep{Barber16}. 

More recent simulations have taken into consideration additional properties of the host galaxy for SMBH seeding. For example, the Horizon-AGN simulation \citep{Volonteri16} required gas and stellar densities and stellar velocity dispersion to exceed certain thresholds for a galaxy to form a black hole, with a seed mass of $10^5\:M_\odot$ adopted. In addition, all the forming black holes needed to be separated by at least 50 comoving kpc, and their formation was only allowed down to $z=1.5$. Adopting similar criteria, the \textsc{obelisk} simulation \citep{Trebitsch21} also applied conditions of gas and stellar density needing to exceed certain thresholds, including Jeans instability of the gas, as well as a required isolation of 50 kpc from other SMBHs to avoid multiple black holes forming in the same galaxy. If all these conditions were satisfied, then a black hole of $3\times10^4 M_\odot$ was assigned to the galaxy. In another approach, the \textsc{romulus} simulation \citep{Tremmel17} employed criteria of a limit on metallicity, a threshold on gas density, and a restricted temperature range for SMBH formation, with a seed mass of $10^6 M_\odot$ adopted. In yet another axample, \cite{Bhowmick22b} have considered a variety of gas-based SMBH seeding prescriptions and a range of seed masses from $\sim 10^4$ to $10^6\:M_\odot$. While the investigation of certain thresholds of physical quantities for SMBH formation is an advance on a simple HMT models, the above studies are still far from being a complete physical description of SMBH formation.

In this work, we focus on a formation scenario in which Population III.1 stars are the progenitors of SMBHs. Pop III.1 stars are defined to be Pop III (i.e., metal free) stars forming in the first dark matter minihalos to form in a given region of the universe and so are isolated from other stellar or SMBH feedback sources \citep{MT08}. In this model it is assumed that in the absence of any significant radiative (or mechanical) feedback, a single dominant protostar forms at the center of the minihalo and has its structure affected by the energy input from Weakly Interacting Massive Particle (WIMP) dark matter self annihilation inside the protostar \citep{SFG08,NTO09,F10,RD15}.
Such protostars maintain relatively cool photospheres and thus low levels of ionizing feedback, which allows efficient accretion of the baryonic content of the minihalo, i.e., $\sim 10^5\:M_\odot$, to form a supermassive star, which subsequently collapses efficiently to a SMBH after a few Myr.

This Pop III.1 seeding mechanism, which is based on locating isolated minihalos, was applied in a cosmological simulation by \citet{Banik2019} (hereafter \citetalias{Banik2019}). The evolution was followed from high redshifts down to $z=10$. The main free parameter in the model is the \textit{isolation distance} ($d_{\rm iso}$), i.e., how far a newly forming minihalo needs to be from previously formed halos in order to be a Pop III.1 source. For a fiducial value of $d_{\rm iso}=100\:$ kpc (proper distance), the model yields co-moving number densities of SMBHs that match the estimated level of the known $z=0$ SMBH population. Note, that in this case (and all other reasonable cases) most minihalos do not form Pop III.1 sources. Rather, most are Pop III.2 sources, which are metal free, but having been disturbed by radiative feedback are expected to undergo significant fragmentation to form only lower-mass (e.g., $\sim 10\:M_\odot$) stars \citep{Greif2006}.

In this paper, we take this Pop III.1 seeding mechanism and extend the results down to the local universe, $z=0$. In \S\ref{sec:methods}, we briefly describe our seeding algorithm and the tools used to apply it. Then we present our results in \S\ref{sec:results}, starting with the evolution of number density of seeded halos down to $z=0$. We compare these results with the HMT scheme, and also discuss the SMBH occupation fraction and clustering properties of seeded halos. Finally, we create synthetic Hubble Ultra Deep Fields (HUDFs) to demonstrate the possibility of using the HUDF to differentiate among different seeding mechanisms. We then present our conclusions in \S\ref{sec:conclusions}.

\section{Methods}
\label{sec:methods}

\subsection{{\sc pinocchio} simulations}

As in Paper I, to test our Pop III.1 seeding mechanism, we used the \textsc{Pinocchio} code \citep{MTT02,Munari17} to generate a cosmological box of 59.7 Mpc (40 $h^{-1}$ Mpc for $h=0.67$) with standard Planck cosmology (Planck Collaboration \citeyear{Planck20}) and study the formation of DM (mini-)halos in that box. {\sc Pinocchio} uses Lagrangian Perturbation Theory \citep[LPT, e.g.,][]{M91} to approximate the evolution of cosmological perturbations in a $\Lambda$CDM universe. For a given set of initial conditions, the code generates outputs in the form of catalogs at different redshifts, which contain mass, position and velocity of the DM halos, and a complete information of the merger histories of all the halos, with continuous time sampling.

This code was written for applications in cosmology, where huge volumes with moderate mass resolution are requested, and its performance heavily depends on the mass resolution adopted. To resolve minihalos of $\sim10^6 M_\odot$ it is necessary to sample a 59.7 Mpc box with $4096^3$ particles; this results in a particle mass of $1.23\times 10^5 M_\odot$, and we adopted a minimum mass of 10 particles (that would be unacceptable for an N-body simulation, but it is acceptable for a semi-analytic code like {\sc Pinocchio}), resulting in a minihalo mass of $1.23 \times 10^6 M_\odot$.

Such a large simulation can only be run on a supercomputer, distributing the computation on a large number of MPI tasks. The construction of halos from collapsed particles is performed in Lagrangian space: the box is divided in sub-boxes, and the grouping algorithm is run on the the particles belonging to its domain. Halos near or across the sub-box borders would not be constructed correctly, so the sub-box is augmented with a "boundary layer" (a ghost region) whose size should scale with the Lagrangian radius $R_{\rm max}$ of the largest halo one expects to find in the simulation volume (that can be of the order of several Mpc). This implies an overhead in memory that can be significant. When dividing a small box into many tasks, the size of the sub-boxes can be of the same order of (if not larger than) $R_{\rm max}$, making the memory overhead unsustainable. The constraint is weakened by stopping the simulation at higher redshift, when $R_{\rm max}$ is still small. As a result, with V4 of {\sc pinocchio} \citep{Munari17} used in Paper I, we were only able to push the simulation down to $z=10$.

We use here the novel V5 of the code, that implements a number of numerical techniques to improve memory efficiency. This code will be presented elsewhere, the strategy to perform halo construction at high resolution is the following. The sub-box is augmented with a boundary layer as large as needed, but instead of storing the properties of all particles in the augmented sub-box we start by storing only the particles that lie in the sub-box (excluding the boundary layer) and are predicted to collapse by $z=0$. Then the halo construction code is run once, collecting a tentative list of halos; after, all the particles that are in the boundary layer and lie within $N_{\rm Lag}$ times the Lagrangian size of any formed halos are added to the list of particles. After collecting the extra information, the halo construction code is run again, generating the final list of halos. Memory occupation thus depends on the parameter $N_{\rm Lag}$; our tests show that $N_{\rm Lag}=3$ guarantees a convergent result, but an extreme run such as the one we present here was possible only by using $N_{\rm Lag}=2$. The 59.7 Mpc box with full $4096^3$ resolution was thus run to $z=0$ on 800 MPI tasks over 100 computing nodes (each with 256 GB of RAM), so the domain was divided into $6\times6\times7.5$ Mpc sub-volumes for halo construction. The resulting halo mass function showed two problems that are presented in greater detail in an Appendix. We discuss here their nature and their implications.

As a consequence of the difficulty of calibrating the formation of halos with a very steep power spectrum, the mass of the first halos is underestimated by a factor of $\sim 2$ at $z\sim30$, decreasing to a negligible value at $z\sim10$. This is a known trend in {\sc pinocchio}, visible, e.g., in Figure 1 of \cite{Munari17} where the $z=3$ halo MF is slightly underestimated in those tests. We are working to improve this prediction, but we do not consider this as a showstopper for several reasons: our seed BHs are already predicted to form very early, so this underestimation only causes us to be slightly conservative in their formation redshift, i.e., in fact they would already have formed at slightly higher $z$. In our simple modeling we are assuming here immediate formation of the protostar and then the SMBH, whereas in reality this might take several Myr or even tens of Myr. The time span that separating $z=32$ from $z=29$ is only $\sim14$ Myr, so neglecting astrophysical timescales leads to an overestimation of formation redshift, which compensates against the underestimation problem. Finally, the minihalo threshold mass can be consider to be a second free parameter of the modeling (although one that has physical motivation to be close to $10^6\:M_\odot$), so one can simply consider our predictions to be valid for minihalo masses of $2.5\times 10^6\ M_\odot$. We add to these arguments the fact that inaccuracies in halo masses do not propagate as inaccuracies in halo positions, that are crucial outcomes of our seeding scheme.

A more serious problem is connected to the inaccurate reconstruction of halos more massive than $10^{12} M_\odot$. Indeed, the small size of the sub-box domain for constructing halos results in a poor reconstruction of massive halos. This problems makes predictions at $z=0$ unreliable. We thus produced the same box at a lower resolution, sampled with $1024^3$ particles, on a single MPI task on a 256 GB node. Again, this was possible thanks to V5 of the code. In this case halo construction is as good as it can be. However, the identification of halos that contain seed SMBHs has been performed in the high resolution box, and though the simulations share the same large-scale structure, matching massive halos in the two boxes is not a clean procedure. We then resorted to this algorithm: starting from the fact that one low-resolution particle contains 64 high-resolution ones, we calculated which particle in the lower resolution box includes the seeded mini-halo, and assigned the seed to the halo that contains that specific low-resolution particle. We checked that results at $z=0$ produced with the low- and high-resolution simulations were consistent, with a significant difference in halo clustering of halos more massive than a certain threshold that is an expected consequence of the inaccurate mass reconstruction and the known relation of halo bias with halo mass. In the following we will present results at $z=0$ based on the low resolution box, unless mentioned otherwise.

\subsection{Seeding scheme}

\begin{figure}
    \centering
    \includegraphics[scale=0.55]{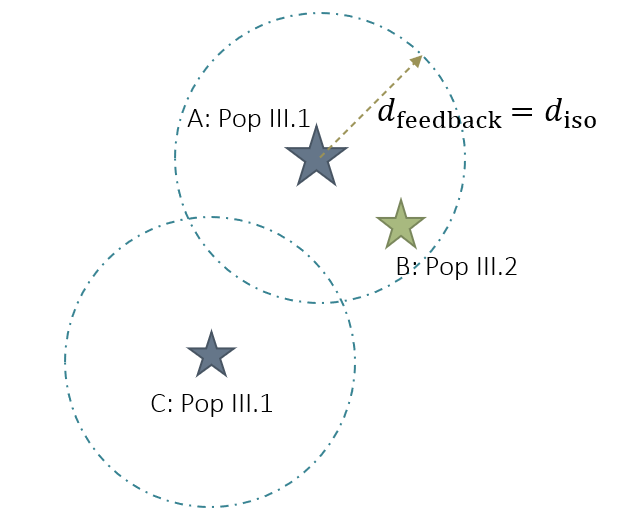}
    \caption{A schematic illustration of the Pop III.1 SMBH seeding scenario depicting the conditions for a star to be isolated enough to be considered as a Pop III.1 star (see text).}
    \label{fig:pop3.1}
\end{figure}

To determine which halos are seeded with a Pop III.1 star and thence SMBH, consider the scenario depicted in Fig. \ref{fig:pop3.1}, unfolding in the early universe. The figure shows three stars A, B and C in different halos where only A and C become Pop III.1 stars whereas B is a Pop III.2 star, depending on the separation and formation order. Star A formed first, which then influenced its environment within a sphere of radius equal to $d_{\rm feedback}$, expected to be primarily radiative feedback. Since this star is in a pristine primordial gas without the influence of any feedback from nearby stars, it is defined to be a Pop III.1 star. Star B, which subsequently forms at a distance less than $d_{\rm feedback}$ from star A, is affected by the feedback and hence is a Pop III.2 star (or even a Pop II star if it has been chemically polluted). Finally, star C forms beyond the regions affected by feedback from sources A and B, and is thus also assigned to be a Pop III.1 star and thus a SMBH. For the model considered here, the feedback distance is set equal to the isolation distance $d_{\rm iso}$. So effectively, the condition for a star to be regarded as a Pop III.1 star is that when it is forming, there should be no previously formed halos present in the sphere of radius $d_{\rm iso}$. We consider $d_{\rm iso}$ as a free parameter in our theory and vary it to match the observed number density of the SMBHs in the local Universe.

\subsection{Seed identification in the dark matter catalogs}

To perform the seed identification analysis from the dark matter catalogs generated by \textsc{pinocchio}, we first divided the entire redshift range (from $z=0$ to the redshift when the first minihalo forms, $z\approx40$) into small bins of widths ranging from $\Delta z = 1, 2$ or $3$, depending on the output catalogs available, which in turn depends on the relative change in positions of (mini)halos. The bins are wider at high redshifts, but smaller at lower redshifts. Then for each redshift interval $(z_l, z_h]$ where ($z_h > z_l$), we utilised \textit{k}-d tree data structure to create a three dimensional map in position space of all the halos existing between $z_h$ and $z_l$. The positions used to create the tree are taken from the output catalog of \textsc{pinocchio} at the lower redshift of the interval ($z_l$). Since the positions are not updated once the tree is constructed, we account for the change in the positions within this redshift interval by finding the maximum change ($\delta$) of position among all the halos existing for the entire redshift range. Then for each minihalo crossing the mass threshold of $10^6 M_\odot$ (or as in the nomenclature of \textsc{pinocchio}: "appearing") at a redshift $z_{\rm app} \in (z_l, z_h]$, we perform a ball search using the \textit{k}-d tree to find all the halos around the appearing minihalo within a sphere of radius $d_{\rm iso} - 2\delta$\footnote{A factor of 2 is multiplied with $\delta$ to account for the change in position of both the minihalo at the center of the sphere and all the other halos within the sphere.}. If there exists even a single halo at the redshift $z_{\rm app}$ within this sphere, then this minihalo is flagged as a halo containing a non-Pop III.1 star at its center. If there are no halos existing at this redshift, then the ball search is performed again with the same minihalo at the center, but this time within a sphere of radius $d_{\rm iso} + 2\delta$. Then for all the halos existing at redshift $z_{\rm app}$ within the shell of radius $d_{\rm iso} \pm 2\delta$, we find the exact distance between the minihalo at the center and all these halos using the exact positions at $z_{\rm app}$. If this distance is greater than $d_{\rm iso}$ for all the halos within the shell, then the minihalo at the center is flagged as a Pop III.1 source, i.e., an SMBH-seeded halo. This process is repeated for each minihalo crossing the threshold mass within the two redshifts, and then this whole procedure is performed again for all the redshift intervals, until the whole redshift range is covered. In this way we are able check the isolation condition for each minihalo appearing in the cosmological box and find all the seeded minihalos. 

At smaller redshifts, the change in positions of the halos ($\delta$) within the redshift intervals becomes comparable to the isolation distance. This implies that the quantity $d_{\rm iso} - 2\delta$ can become negative (in our simulation box, this happens at around $z\approx15$ for $d_{\rm iso} = 50$ kpc). In this case, the ball search is directly performed in a sphere of radius $d_{\rm iso} + 2\delta$, and then the exact distances between the minihalo at the center and all the other halos existing at $z_{\rm app}$ is calculated.

This division of the entire redshift interval and creating the \textit{k}-d only at specific redshifts is performed to avoid reconstructing the tree with the up-to-date position at every instance a new minihalo appears. Since the number of minihalos is very large, it becomes highly expensive computationally to reconstruct the tree with updated positions each time a new minihalo appears.

\section{Results}
\label{sec:results}

\subsection{Number density evolution}
\label{sec:nd_evol}

\begin{figure*}
    \centering
    \includegraphics[scale=0.5]{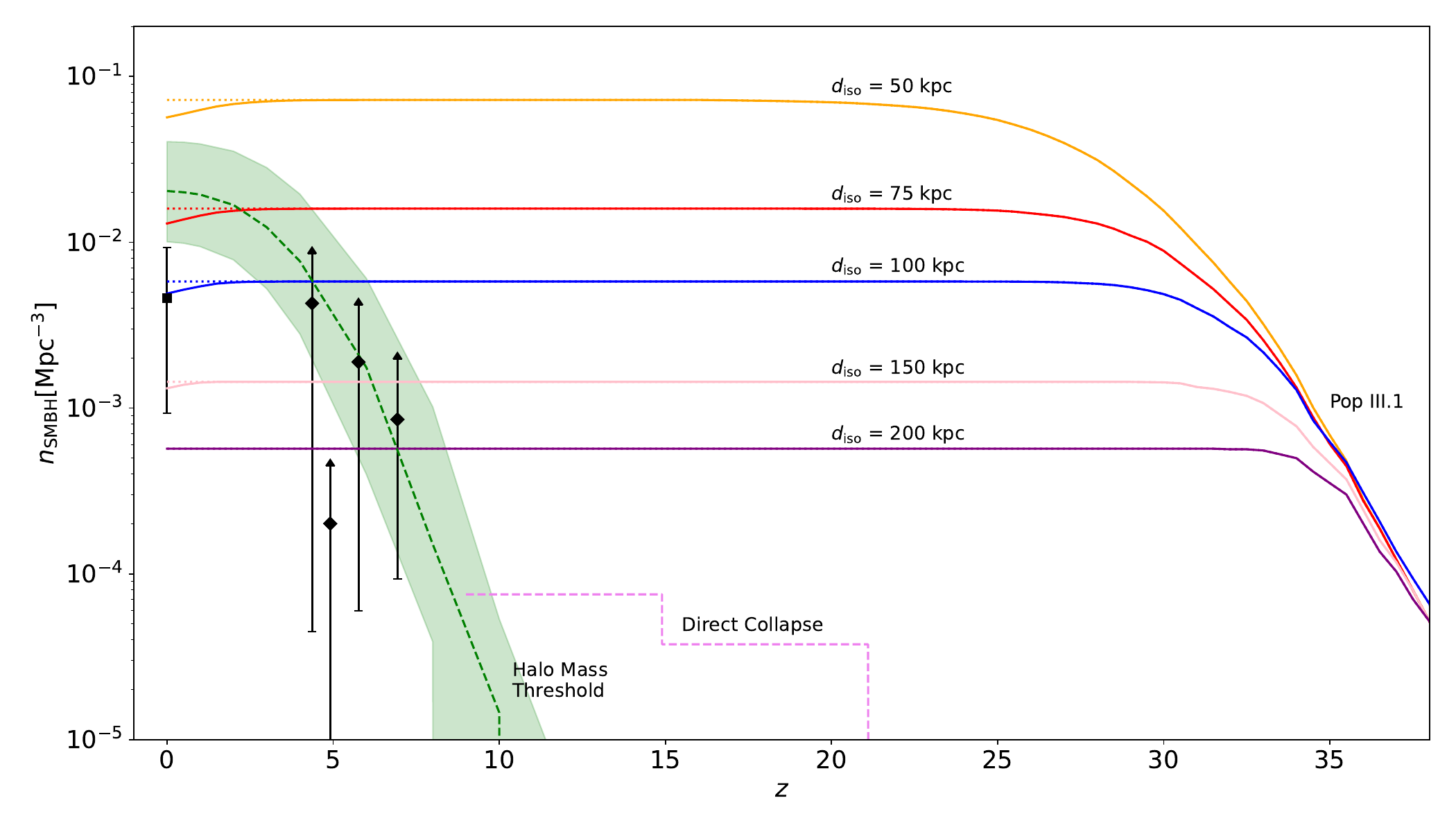}
    \caption{Evolution of the comoving number density of SMBHs, $n_{\rm SMBH}$, for different theoretical models. Results for Pop III.1 models with several values of isolation distance (in proper distance) are shown, as labelled. The dotted lines show the total number of SMBHs that ever formed, while the solid lines show remaining number of seeded halos after accounting for mergers. An example Halo Mass Threshold (HMT) model is shown by the dashed green line in which each halo with mass higher than $m_{\rm th}=7.1\times10^{10}M_\odot$ is seeded (see text). The green shaded region shows the effect of lowering and raising $m_{\rm th}$ by a factor of 2. The violet dashed line shows the results of a simulation modeling SMBH formation via direct collapse \citep{Chon16}. The black solid square indicates an estimate for the number density of SMBHs at $z=0$ assuming each galaxy with luminosity higher than $L_{\rm min} = 0.33 L_*$ contains one SMBH, with the range shown by the error bar obtained by varying $L_{\rm min}$ from $0.1L_*$ to $L_*$. The black diamonds are estimated lower limits of $n_{\rm SMBH}$ from JWST observations of Type I AGN \citep{Harikane23}.}
    \label{fig:nd}
\end{figure*}

As explained in the last section and in \citetalias{Banik2019}, we identify SMBH-seeded halos by the condition that the isolation sphere of radius $d_{\rm iso}$ around a newly forming minihalo is not populated by any other existing halo (of mass greater than our minihalo threshold mass). The obtained results for the evolution of number density for different values of $d_{\rm iso}$ (in proper distance units) are shown in Fig. \ref{fig:nd}. The colored dotted lines show the number density evolution of total number of SMBHs, whereas the colored solid lines show the number density for seeded halos (which are slightly smaller, especially at lower redshifts, due to mergers). Compared to the number densities in Figure 1 of \citetalias{Banik2019}, the values obtained here are moderately lower (by a factor of $1.45$ for 100 kpc and $1.65$ for 50 kpc) because we have considered periodic boundary conditions when identifying the seeds, which was not done in \citetalias{Banik2019}.

Fig. \ref{fig:nd} also shows some observational estimates of $n_{\rm SMBH}$. An estimate at $z=0$, presented in \citetalias{Banik2019}, is calculated by assuming that each galaxy with luminosity greater than $L_{\rm min}=0.33L_*$ hosts a SMBH, with the error bar around this point assuming a range of $L_{\rm min}$ from 0.1 to 1.0~$L_*$. Note, $L_*$ is the characteristic luminosity corresponding to $M_{B} = -19.7+5\log{h} = -20.55$ \citep[e.g.,][]{N02}. Recent observations of high redshift AGNs from JWST surveys have started providing lower limits on the number density of SMBHs in the early universe, with one such estimate presented by \cite{Harikane23}, from the sample of \cite{Nakajima23} (black diamonds in Fig. \ref{fig:nd}). Their estimate provides lower bounds on the observed number density of Type I AGN at redshifts $z=4$ to 7.

From Fig. \ref{fig:nd} illustrates the expected behaviour that as the isolation distance is reduced, the number of formed SMBHs increases, i.e., it is easier to satisfy the isolation distance criterion. We can also conclude that for a certain range of $d_{\rm iso}$ ($\approx 90$ kpc to $170$ kpc), the number density obtained is in reasonable agreement with the $z=0$ estimate. Thus, the case with $d_{\rm iso}=200\:$ kpc is disfavoured simply by its inability to produce enough SMBHs. A key feature of the fiducial model, i.e., with $d_{\rm iso}=100\:$kpc, is that {\it all SMBHs have formed very early in the Universe: the process is essentially complete by $z\simeq 25$.} 

Fig. \ref{fig:nd} also shows results for an example halo mass threshold (HMT) model (shown by green dashed line) in which each halo more massive than $m_{\rm th}=7.1\times10^{10}M_\odot$ is seeded \citep[e.g., the Illustris project:][etc.]{Vogelsberger14, Sijaki15}; note, this seeding scheme is driven by the mass resolution of the simulation, i.e., halos are seeded as soon as they are resolved with a sufficient number of particles). The main difference compared to the fiducial Pop III.1 model is in the overall number density of SMBHs at $z\gtrsim 5$. 

We also show the results of a simulation by \citet{Chon16} modeling the formation of SMBHs via the direct collapse mechanism. Here they simulated a $20 h^{-1}$Mpc box and found only two SMBHs formed (at $z\simeq15$ and 21). Even though this simulation was only run down to $z=9$, the number density is not expected to increase much at lower redshifts, given the conditions assumed to be needed for direct collapse, i.e., massive, irradiated, tidally-stable, metal-free halos. While this model allows some SMBHs to form relatively early, as discussed in \S\ref{sec:introduction}, the overall number densities achieved by this mechanism are much smaller than are needed to explain the entire observed SMBH population.

We quantify the number of mergers that occur between seeded halos in the Pop III.1 models. Table \ref{table:seed_dist} shows the total number of SMBHs that formed ($N_{\rm SMBH,form}$) and the number of halos containing them at $z=0$ ($N_{\rm SMBH}(z=0)$). Assuming efficient merging of SMBHs that are in the same halo, then the number of mergers is $\Delta N_{\rm SMBH}=N_{\rm SMBH,form}-N_{\rm SMBH}(z=0)$. A feature of the Pop III.1 seeding mechanism is that SMBHs are initially spread out from each other, so that there are relatively few binary SMBHs and few mergers. A detailed analysis of the mergers including the binary (and higher order multiples) AGN number densities, and the gravitational wave background emanating from these mergers will be discussed in a future paper in this series.

\begin{table}
\caption{Total number of formed SMBHs ($N_{\rm SMBH,form}$), total number of SMBHs remaining at $z=0$ assuming efficient mergers ($N_{\rm SMBH}(z=0)$), the difference between these ($\Delta N_{\rm SMBH}=N_{\rm SMBH,form}-N_{\rm SMBH}(z=0)$), which is equivalent to the number of mergers, and the fraction of original SMBHs that are destroyed by mergers ($f_{\rm merger}=\Delta N_{\rm SMBH}/N_{\rm SMBH,form}$).}
\centering
\begin{tabular}{@{}ccccc@{}}
\toprule
$d_\text{iso}$ {[}kpc{]} & $N_{\rm SMBH,form}$ & $N_{\rm SMBH}(z=0)$ & $\Delta N_{\rm SMBH}$ & $f_{\rm merger}(\%)$   \\ \midrule
50                       & 15356     & 12051     & 3305        & 21.52 \\
75                       & 3394      & 2760      & 634         & 18.68 \\
100                      & 1234      & 1043      & 191         & 15.48 \\
150                      & 306       & 280       & 26          & 8.50  \\ 
200                      & 121       & 116       & 5           & 4.13  \\
\bottomrule
\end{tabular}
\label{table:seed_dist}
\end{table}

A caveat of our seeding model is that at small redshifts, around $\lesssim6$, the isolation distance in comoving units becomes so small that many minihalos that appear after this redshift start satisfying the isolation criteria. This effect would result in an increase in number density by around 2 orders of magnitude by $z=0$ from the converged values around $z\approx20$, for all cases of $d_{\rm iso}$. However, since reionization has completed by $z\approx8$ (Planck Collaboration \citeyear{Planck20}), we assume that the formation of Pop III.1 sources is also not possible below this redshift. Hence, in our analysis, we set a limit of seed formation to be only possible until $z=8$. For most cases of the isolation distances we considered ($\geq 75$ kpc), the number density is already converged at redshifts greater than $z=20$. However, for the case of 50 kpc, new seeds still keep on appearing until $z=8$ (although below $z=15$ the total number only increases by about 1\%).

\begin{figure*}
    \centering
    \includegraphics[scale=0.39]{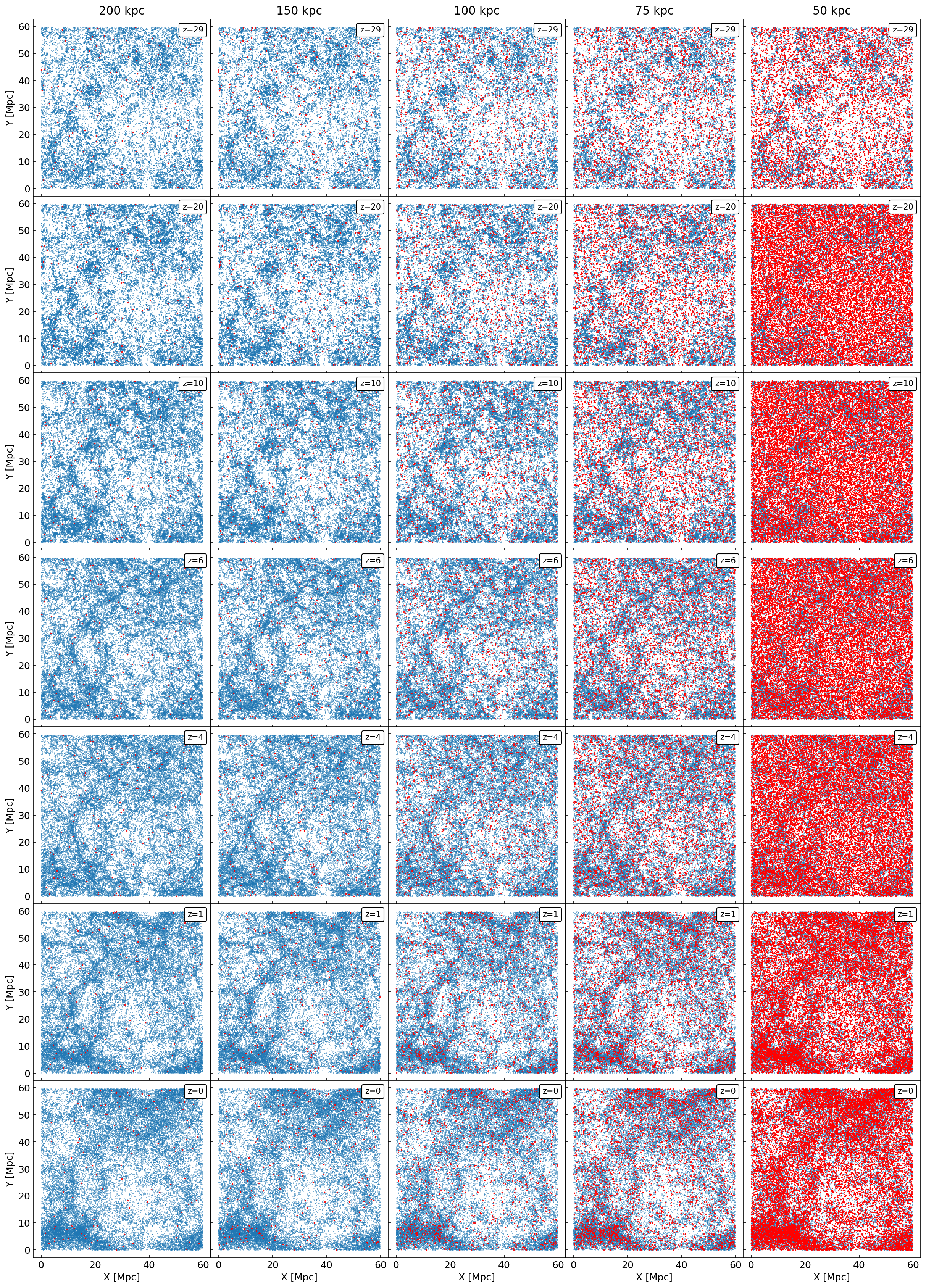}
    \caption{Projection of the positions of seeded halos (\textit{red}) and non-seeded halos (\textit{blue}) in the XY plane of the box for different isolation distances. The redshift is shown in the top right corner of each panel (same for each row). Only the 30,000 most massive non-seeded halos within each panel are shown for ease of visualisation.}
    \label{fig:box_seed_evolution}
\end{figure*}

In Figure \ref{fig:box_seed_evolution}, we show a visual representation of the seeded halos in the box at different redshifts, for all the isolation distances considered in Fig. \ref{fig:nd}. As discussed, the 50 kpc case is the most crowded with the highest number of seeded halos at every epoch shown. Initially all the seeds emerge in a relatively unclustered manner, but eventually the clustering increases as lower-mass seeded halos migrate towards more massive halos and merge with them in overdense regions. We perform a more detailed analysis of clustering in \S\ref{sec:clustering}.

\subsection{Occupation fraction of seeded halos}
\label{sec:occ_frac}

\begin{figure*}
    \centering
    \includegraphics[scale=0.51]{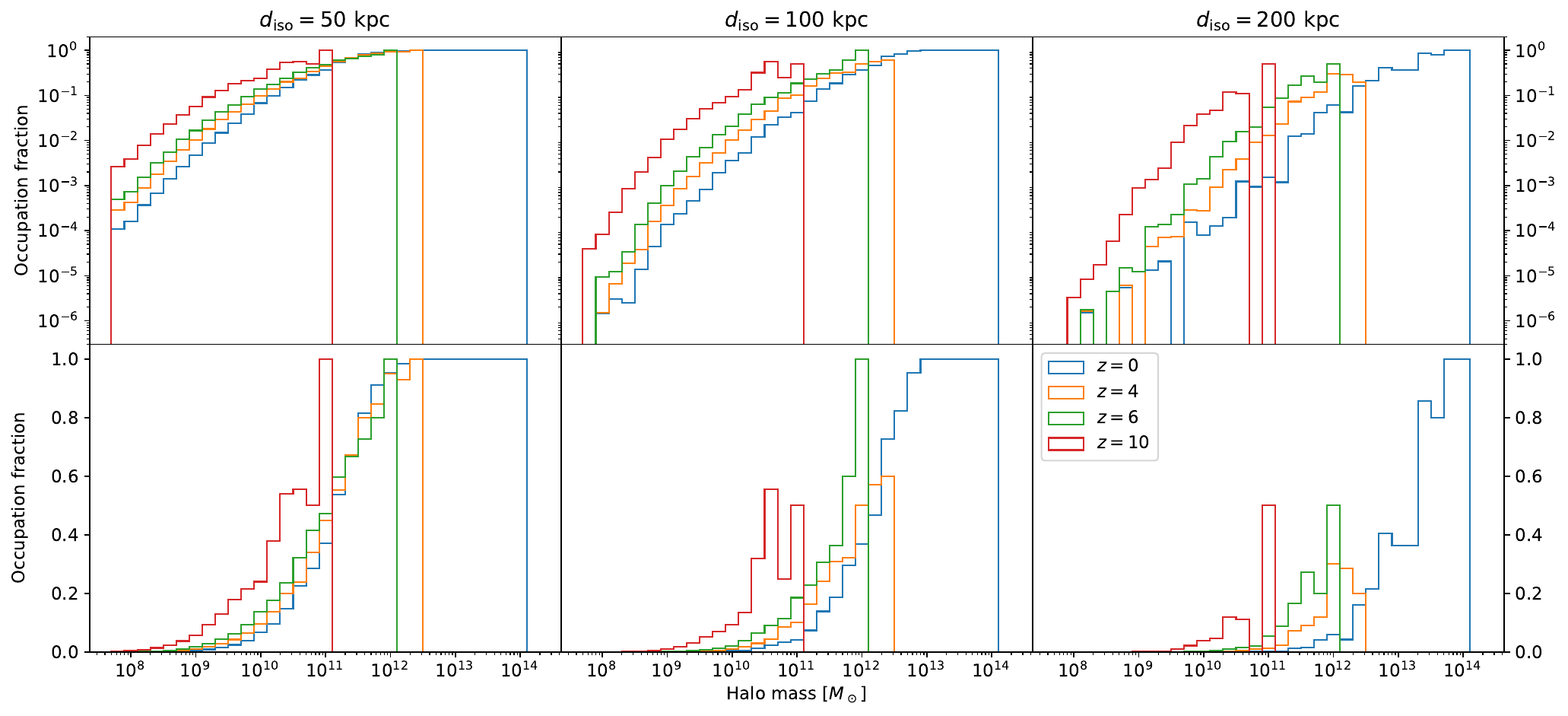}
    \caption{Evolution of SMBH occupation fraction of halos for different cases of $d_{\rm iso}$. Top row depicts the fraction in log scale, while the bottom row shows the same data in linear scale. The mass bins are divided into equal bins of width 0.2 dex.}
    \label{fig:occ_frac}
\end{figure*}

From observations of local galaxies, it appears that almost all massive galaxies contain a nuclear SMBH. This implies that the SMBH occupation fraction of halos should approach unity as halo mass rises. Figure \ref{fig:occ_frac} shows the evolution of occupation fraction from one realization of our 59.7 Mpc box, through 4 different redshifts for halos ranging from \textbf{$[5\times10^7, 2\times10^{14}] M_\odot$} (the upper limit of the mass range is chosen to include the most massive halo at $z=0$ in our $1024^3$ resolution simulation box, measuring \textbf{$1.2\times10^{14} M_\odot$}). As expected, with the decrease in the isolation distance, more and more halos are seeded and hence the occupation fraction is higher compared to the same mass range for larger $d_{\rm iso}$. All the fractions at $z=0$ approach unity for the most massive halos, independent of the isolation distance. Interestingly, the most massive halo is not always occupied by a SMBH throughout the redshift evolution in our simulations. For example, at $z=4$ there can be significant fractions of the most massive halos, i.e., $\sim 10^{12}\:M_\odot$, that are not seeded, as in the case of $d_{\rm iso}=$100 kpc. Figure \ref{fig:occ_frac} also shows that for $d_{\rm iso}=200\:$kpc the occupation fraction for halos with masses $\sim 10^{12}\:M_\odot$ at $z=0$ is quite small, $\lesssim 0.1$, which is a further indication that it produces too few SMBHs.

\begin{figure*}
    \centering
    \includegraphics[scale=0.53]{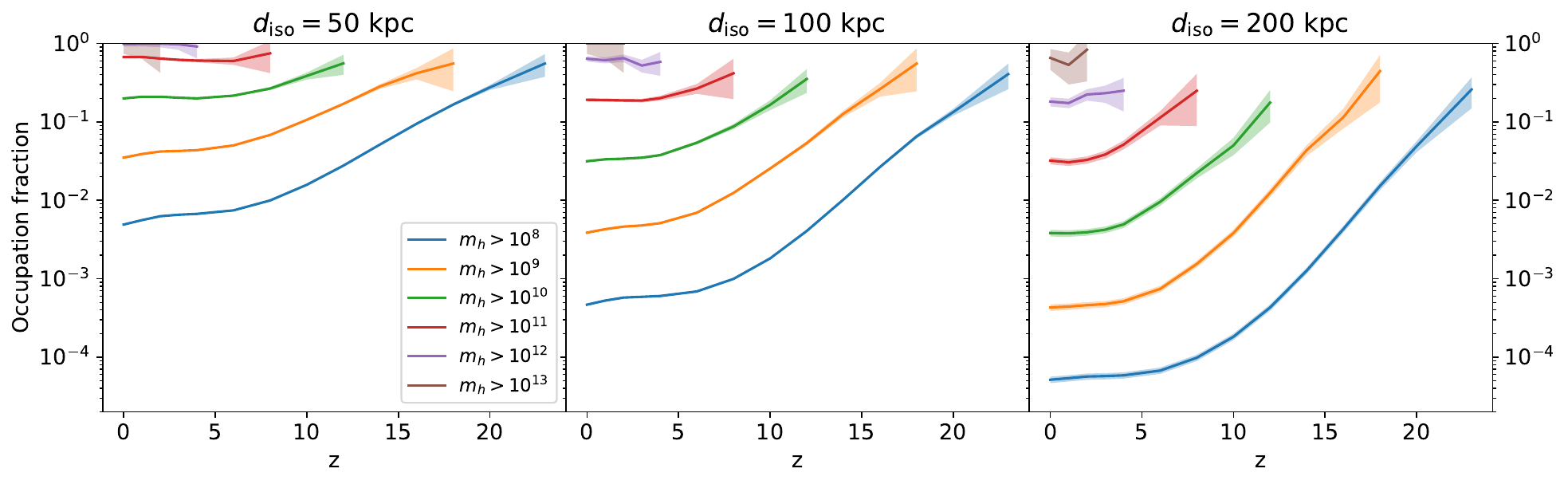}
    \caption{Cumulative occupation fractions of halos having masses greater than a given value (see legend). The shaded region represents $\pm1\sigma$ error due to counting statistics.}
    \label{fig:occ_frac_trend}
\end{figure*}

Figure \ref{fig:occ_frac_trend} shows the evolution of the cumulative occupation fraction, i.e., for all halos more massive than $\{10^8, 10^9, 10^{10}, 10^{11}, 10^{12}, 10^{13}\} M_\odot$, for three different cases of isolation distance. If we consider only the most massive halos ($>10^{13} M_\odot$), the fraction is close to one (as also evident from Fig. \ref{fig:occ_frac}). At a given redshift, as we consider less massive halos, the occupation fraction decreases. At a given mass threshold, as we move out to higher redshift the occupation generally rises, since these halos become relatively more extreme members of the global halo population. Interestingly, the occupation fraction for all halos more massive than $10^8$ and $10^9 M_\odot$ ($10^{10} M_\odot$ as well, although to a lower degree) at $z=0$ differ by factors of approximately 10 among the three cases of isolation distances considered, reflecting the same differences in the global number densities at $z=0$ (see Fig. \ref{fig:nd}).

\begin{figure*}
    \centering
    \includegraphics[trim=0.25cm 0 0.25cm 0,clip,scale=0.51]{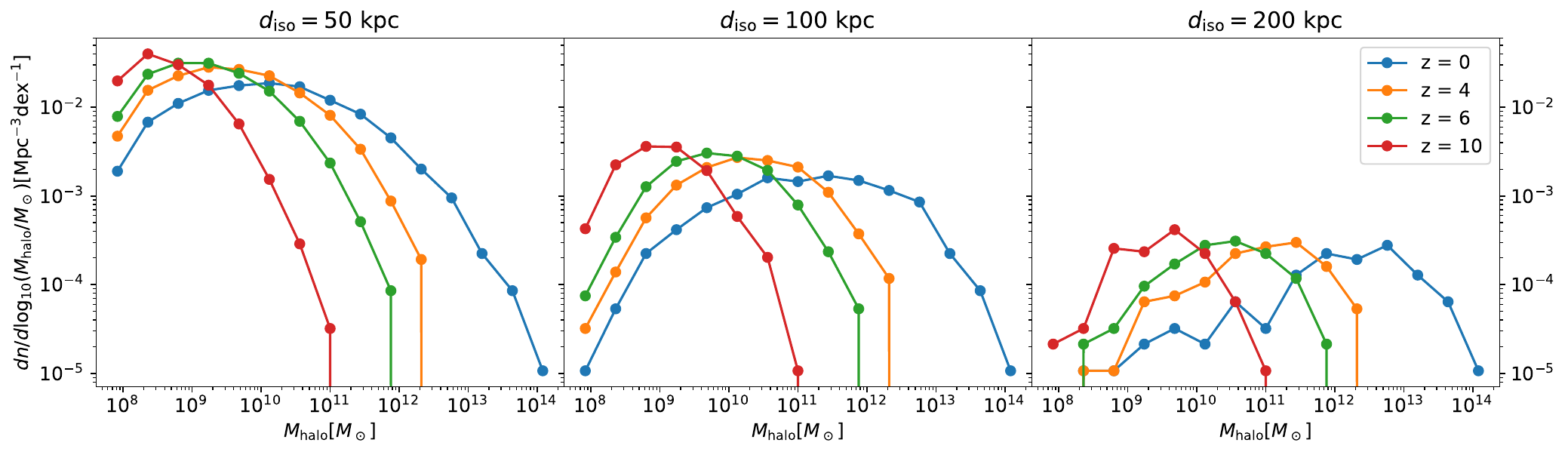}
    \caption{Mass function of seeded halos at different redshifts for $d_{\rm iso}=\:$50, 100 and 200 kpc cases (left to right).}
    \label{fig:mass_fun_seeded}
\end{figure*}

To obtain a better understanding of the mass function of the seeded halos, in Figure \ref{fig:mass_fun_seeded} we present the distribution functions of these halos for the $d_{\rm iso}=50$, 100 and 200~kpc cases, including their evolution with redshift. We see that, as expected, these mass functions evolve to higher masses as the universe evolves from $z=10$ down to $z=0$. The peak of the seeded halo mass function is lower for smaller values of $d_{\rm iso}$. However, the distributions are quite broad, indicating significant fractions of SMBHs in relatively low-mass halos, even at $z=0$. In a future paper in this series, these seeded halo mass functions and the properties of their host galaxies will be compared to SMBH census data, especially focusing on properties derived in the local universe.

\subsection{Clustering}
\label{sec:clustering}

\begin{figure*}
    \centering
    \includegraphics[trim=0.5cm 0 0.5cm 0,clip,scale=0.48]{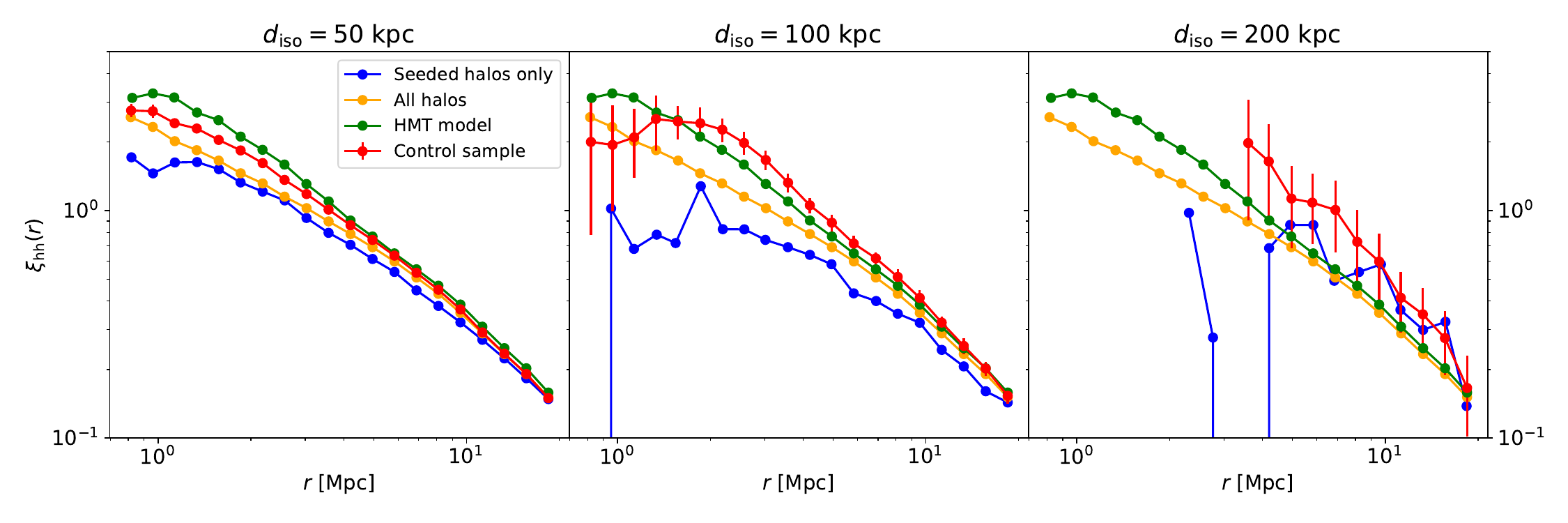}
    \caption{The 3D 2 point correlation function for the seeded halos more massive than $10^{10} M_\odot$, at $z=0$ for different isolation distances. The blue points show the correlation function for only the halos containing SMBHs, while the orange points show the correlation for all the halos, with or without a SMBH. For the red points, we randomly select halos from the pool of all the halos, but with the same number and mass distribution as the seeded halos. The error bars indicate $1\sigma$ deviations from the mean value from randomly sampling 50 times. The green points show the correlation for halos seeded according to the halo mass threshold (HMT) scheme, in which all the halos greater than $m_{\rm th}=7.1\times 10^{10}M_\odot$ are seeded.}
    \label{fig:clust}
\end{figure*}

We perform a clustering analysis using the \textsc{corrfunc} library \citep{Sinha20} for \textsc{python}, and the results are shown in Fig. \ref{fig:clust}. By sampling $r$ in 20 logarithmic bins of $r_{\rm min} = 0.5$ Mpc/h to $r_{\rm max}=13.3$ Mpc/h, we evaluate the 3D 2-point correlation function\footnote{All the correlation functions presented in this section have been corrected by analytically adding large scale clustering modes corresponding to scales larger than the box size. Refer to appendix \ref{appendix:ls_corr} for more details.} (2pcf) $\xi_{\rm hh}(r)$ for all halos more massive than $10^{10} M_\odot$ at $z=0$. Since \textsc{pinocchio} only evolves dark matter halos, the information of substructures such as subhalos within halos is not stored or tracked. This implies that only radial scales greater than the size of a typical dark matter halo (3 to 4 Mpc at $z=0$), are relevant for consideration. In other words, the correlation function presented here does not include the one-halo term. From the figure, we observe that the clustering of the SMBH-seeded halos (blue points) is always lower compared to other cases. This is expected because of the nature of our model, which results in larger distances between SMBHs and hence smaller clustering amplitude. The plots for $d_{\rm iso}=$ 50 and 100 kpc clearly depict this, while the case of 200 kpc suffers from low number statistics. The red points, which represent the clustering of random halos with the same number and mass distribution as of the seeded halos, are generally more than $1\sigma$ higher than the blue points, except at the largest scales. This can be clearly seen for the fiducial case of 100 kpc. We also show the clustering for the fiducial case of HMT schemes with $m_{\rm th}=7.1\times10^{10}M_\odot$ \citep{Sijaki15}, depicted by green points. This model also generally shows higher clustering than our Pop III.1 seeding model. Thus a clustering analysis of census of a local Universe ($z=0$) survey of all (or a significant fraction) of SMBHs has the potential to distinguish between these SMBH seeding mechanisms.

\begin{figure*}
    \centering
    \includegraphics[scale=0.6]{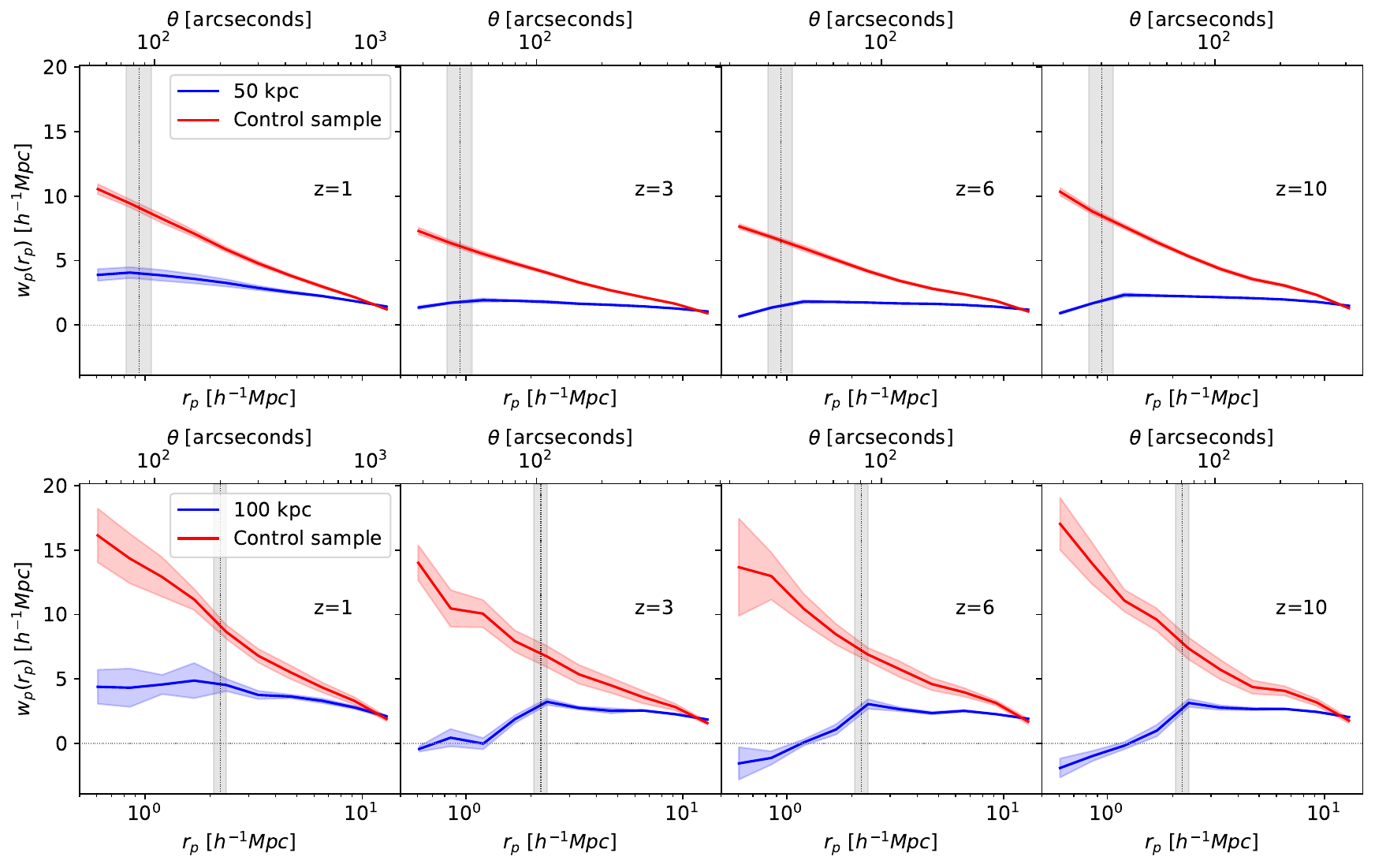}
    \caption{Evolution of projected correlation function for $d_{\rm iso}=$ 50 kpc (top row) and 100 kpc (bottom row) cases. The blue line is the average after computing the correlation of the seeds from 3 orthogonal sides of the box and the shaded region represents the $1\sigma$ spread. The control sample is the correlation of halos selected randomly but with the same mass and number distribution as the seeded halos at that redshift. The red line refers to the average after randomly sampling 10 times and the shaded region refers to $1\sigma$ deviations from the mean. The vertical grey line refers to the size of the isolation radius at the mean formation redshift ($d_{\rm iso}(\bar{z}_{\rm form})$) of the seeded halos, and the grey region represents $1\sigma$ deviation from the mean. For 100 kpc, $\bar{z}_{\rm form}=32.08$, and for 50 kpc, $\bar{z}_{\rm form}=27.14$. The angular axis on top of each panel corresponds to the angular scale of $r_p$ projected on the sky at the respective redshift.}
    \label{fig:clus_evolution}
\end{figure*}

In Figure \ref{fig:clus_evolution}, we show the evolution of the projected correlation function for the $d_{\rm iso}=$50 and 100 kpc cases (blue lines), compared to halos with the same mass and number distribution as the respective seeded halos (red lines). As seen in the 3D 2pcf, the clustering of the seeded halos is always lower than the randomly selected halos and this trend is observed even at higher redshifts. Furthermore, there is a significant drop of the clustering amplitude of the seeded halos for scales lower than $d_{\rm iso}(\bar{z}_{\rm form})$ (vertical grey band), a signature of feedback cleared bubbles, first discussed in \citetalias{Banik2019} for $z\geq 10$. Here we see that this signature of suppressed clustering persists to lower redshift, although is gradually diminished as the Universe evolves to a more clustered state.

We emphasise that comparing our clustering predictions at redshifts greater than 1 or 2 is not feasible with currently available observational data. The measurements from a range of luminosity of AGNs at these redshifts imply minimum halo masses of $\sim5\times10^{11} h^{-1}M_\odot$ at $z\sim3$ \citep{Allevato14} to more than $10^{12}h^{-1}M_\odot$ at $z\sim4$ \citep{He18}. For our 59.7 Mpc box, the number of seeded halos above these thresholds are quite low. For instance, for the $d_{\rm iso}=$100 kpc case, only around 6\% of sources are above this threshold at $z=3$ and only $0.7\%$ sources are more massive than $10^{12}h^{-1}M_\odot$ at $z=4$. If we apply these halo mass cuts on our seeded halos, then the clustering signal is too noisy to make any decent comparison with the observational data. Moreover, at high halo masses the occupation fraction approaches unity, so for the measured clustering of bright AGNs, hosted in relatively massive halos, we expect that they may cluster as their host halos, with no appreciable difference with respect to currently used models. More data on AGN, especially those that are present in lower-mass halos/galaxies is needed to test the models.

\begin{figure}
    \centering
    \includegraphics[scale=0.43]{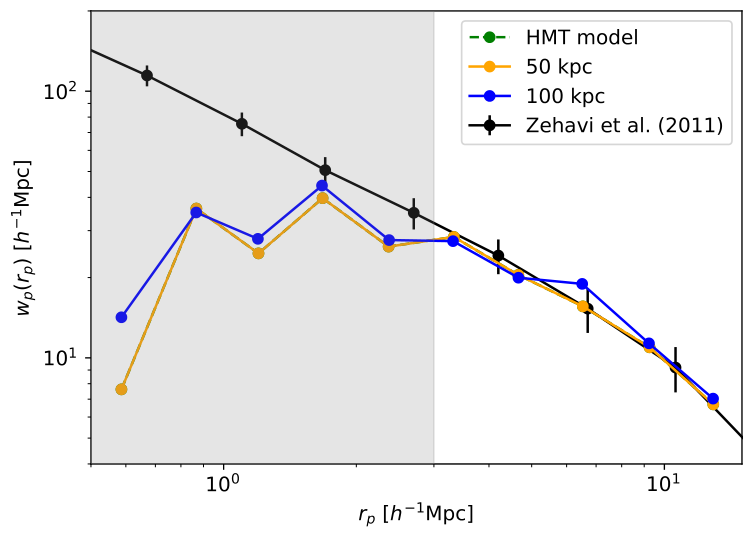}
    \caption{Comparison of the results for the projected correlation function $w_p(r_p)$ obtained from our simulations for $d_{\rm iso}=$50 kpc, 100 kpc and the HMT scheme at $z=0$ with the observational data from \citet{Zehavi11} for a $M_{r}<-19.0$ magnitude cut. The shaded region shows scales smaller than the size of a typical halo at $z=0$, i.e., $r_p<3 h^{-1}$Mpc, which are not of interest for our comparison due to limitations of our model (lack of sub-halos). The HMT scheme and 50 kpc models overlap, as all halos above the threshold are seeded for that value of $d_{\rm iso}$.}
    \label{fig:wp_zehavi}
\end{figure}

As a crude comparison, in Figure~\ref{fig:wp_zehavi} we include the clustering measurements from \citet{Zehavi11}, who performed the projected clustering analysis of volume-limited sample of 570,000 galaxies from the Seventh Data Release \citep{Abazajian09} of the Sloan Digital Sky Survey \citep[SDSS,][]{York00}. The galaxies used in their data extend out to $z=0.25$, with a median redshift of $z\sim0.1$. We compare our results at $z=0$ for $d_{\rm iso}=$50 and 100 kpc, along with the HMT scheme, with their galaxy luminosity threshold cut result for $M_r<-19.0$. We computed the relation between DM halo mass and $r$-band absolute magnitude by comparing the clustering amplitude of \textsc{pinocchio} DM halos with Zehavi et al.'s measurements, minimising the $\chi^2$ of the clustering amplitude only for $r_p>3 h^{-1}$~Mpc (to avoid the one-halo clustering scales); for $M_r<-19.0$ we find a clustering-matched halo mass of $M_{\rm PIN}^{-19.0}=1.91\times10^{12}h^{-1}M_\odot$, higher than the value suggested in that paper ($M_{\rm zehavi}^{-19.0}=2.55\times10^{11} h^{-1}M_\odot$); this is not surprising, given the different cosmology assumed in 2011. we then applied this halo mass cut on our $d_{\rm iso}=$ 50 and 100 kpc sources, as well as the HMT scheme, and compared the projected correlation function for the $M_r<-19.0$ threshold galaxies in Figure~\ref{fig:wp_zehavi}. For the region of interest, the clustering of the seeded halos shows good agreement, within the errors, with the observations. The $d_{\rm iso}=50$ kpc correlation completely overlaps the HMT one because all the sources more massive than $M_{\rm PIN}^{-19.0}$ are seeded in this model. Also, at this high-mass cut, most of the $d_{\rm iso}=$ 50 kpc sources are also seeded in the $d_{\rm iso}=$ 100 kpc model, and hence their clustering follows similar trends. This is due to the fact that the occupation fraction approaches unity for the most massive halos (see \S\ref{sec:occ_frac}) for all the isolation distances, and since the mass cut is high, this means that most, if not all, the halos are seeded, regardless of the isolation distance.

\subsection{Ultra Deep Field}

\begin{figure*}
    \centering
    \begin{subfigure}[b]{0.49\textwidth}
        \includegraphics[width=\textwidth]{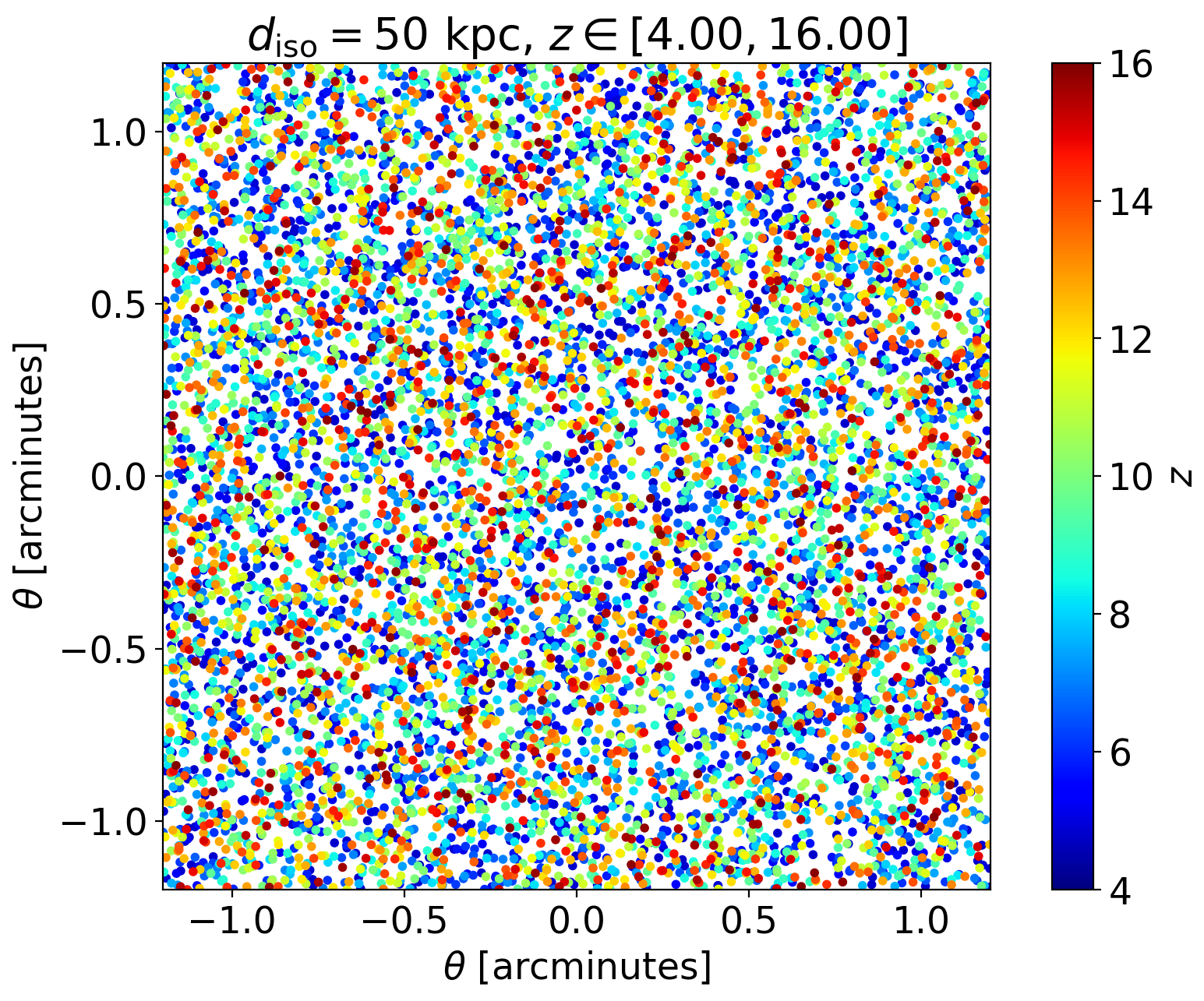}
        \caption{50 kpc}
        \label{fig:udf_50}
    \end{subfigure}
    \begin{subfigure}[b]{0.49\textwidth}
        \includegraphics[width=\textwidth]{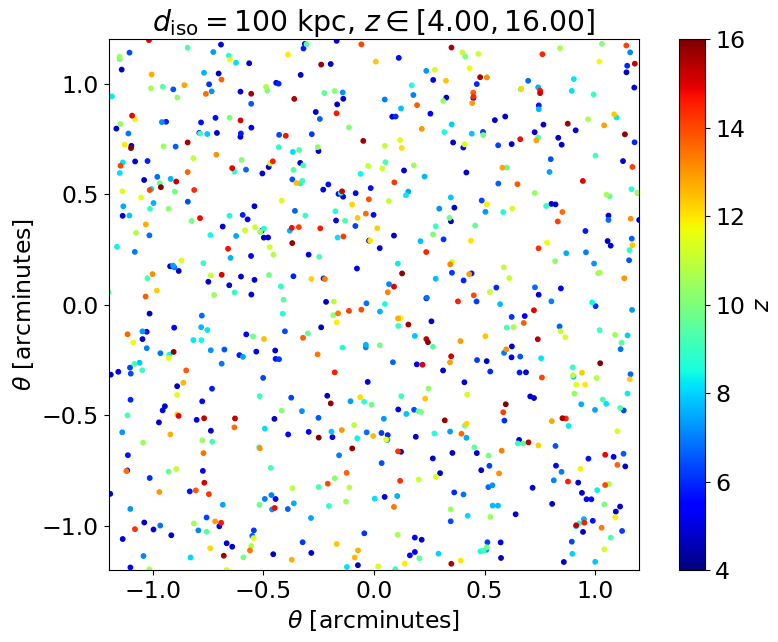}
        \caption{100 kpc}
        \label{fig:udf_100}
    \end{subfigure}
    \caption{Synthetic Hubble Ultra Deep Field (HUDF) consisting of only the seeded halos for $d_{\rm iso} = 50$ kpc and 100 kpc cases over a redshift range from 4 to 16.}
    \label{fig:udf}
\end{figure*}

One potential way to compare our model with observational data is to count the number of SMBHs (i.e., appearing as AGN) present in projected deep fields of the Universe, such as the Hubble Ultra Deep Field \citep[HUDF,][]{Beckwith06,Ellis13}. We thus create a synthetic ultra deep field (UDF) populated with SMBHs that have formed in our simulations. To achieve this, we use snapshots of halos at different redshifts in the 59.7 Mpc cosmological box, using the highest resolution run. We pierce the box orthogonally from random positions (avoiding repetitions) and then stack the fields in redshift space to generate the light cone of a 2.4 arcminute side length (i.e., same as the HUDF). Figure \ref{fig:udf} shows our constructed HUDF, for $d_{\rm iso}=$ 50 kpc and 100 kpc. The fields shown are for the redshift range $z\in[4,16]$, with the number of halos in the field equal to 9352 and 764 for $d_{\rm iso}= 50$ kpc and 100 kpc, respectively. As expected, the field for the 50 kpc case is much more densely populated with seeded halos as compared to 100 kpc.

\begin{figure*}
    \centering
    \includegraphics[scale=0.58]{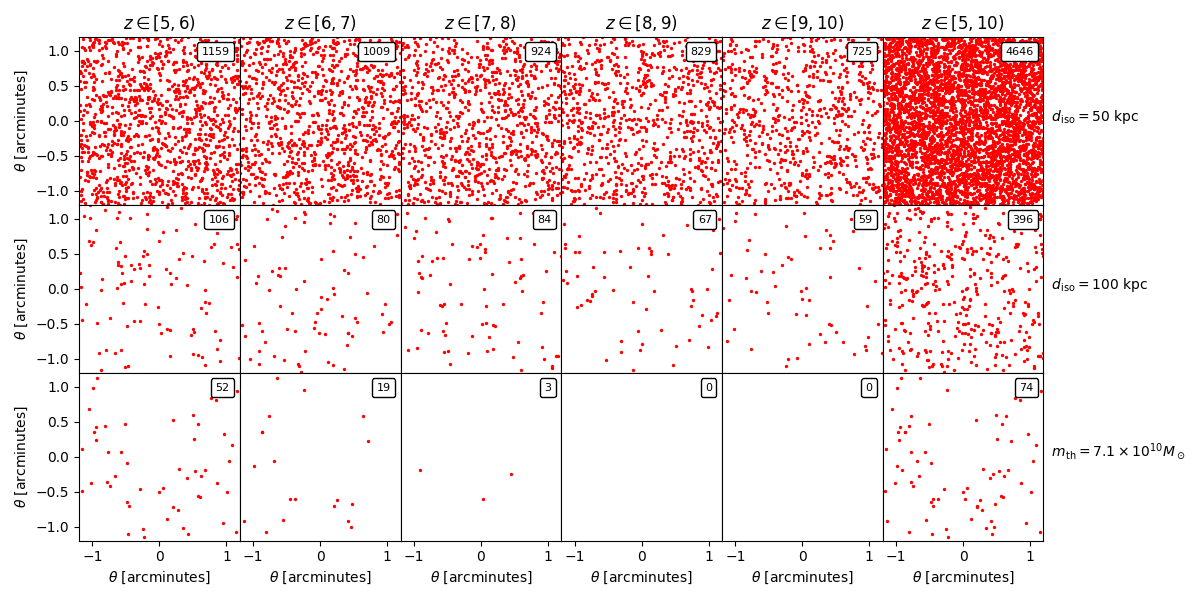}
    \caption{The distribution of SMBHs in  redshift intervals in the range $z=5-10$ in a synthetic HUDF, where the last column shows all the sources. The first row shows the case for $d_{\rm iso}=$50~kpc. The second row shows the case for $d_{\rm iso}=$100~kpc. The third row shows the distribution from the fiducial HMT scheme with $m_{\rm th}=7.1\times10^{10} M_\odot$. The total number of SMBHs in each panel are indicated in the top right corners of each. 
    }
    \label{fig:udf_grid}
\end{figure*}

Figure \ref{fig:udf_grid} shows the distribution of SMBHs within the redshift range $z=5-10$ in our synthetic HUDF, where we also display the number of sources in redshift bins of $\Delta z=1$. The total number of sources in the field (\textit{last column}) for the fiducial $d_{\rm iso}=$100 kpc model is five times higher than the fiducial HMT scheme. Thus a census of AGNs at high redshifts ($z\gtrsim7$) can distinguish between these models. Since the number density of sources in the HMT scheme is quite low (effectively 0 for redshifts $\gtrsim8$ or 9), finding even a handful of sources at these redshifts can put stringent constrains on this seeding scheme. In Table~\ref{table:light_cone}, we show the number of seeds in the field for an extended redshift range by averaging from multiple random realisations of the light cone, and by integrating the number density over the field volume. Almost all the averages in the redshift bins from the light cone are within $1\sigma$ of the analytically calculated value from the number density. The analytic numbers also show the drastic difference in the number of sources in the different seeding schemes at high redshifts.

\begin{table*}
\caption{Number of SMBHs in our synthetic HUDF, calculated by averaging over 100 random realizations of the light cone (\textit{From light cone} column) and by integrating the global number density (\textit{From number density} column) over the redshift ranges, for $d_{\rm iso}=100$ kpc and the fiducial HMT scheme with $m_{\rm th}=7.1\times10^{10}M_\odot$. The errors on the averaged values correspond to $1\sigma$ deviations. Note that all the numbers are rounded to the nearest integer.}
\centering
\begin{tabular}{@{}ccccc@{}}
\toprule
z range & \multicolumn{2}{c}{100 kpc}           & \multicolumn{2}{c}{HMT} \\ 
        & From light cone & From number density & From light cone           & From number density          \\ \midrule
4-5     & $110\pm8$       & 101                 & $86\pm19$                 & 105                          \\
5-6     & $92\pm6$        & 90                  & $36\pm10$                 & 49                           \\
6-7     & $85\pm5$        & 81                  & $13\pm6$                  & 18                           \\
7-8     & $74\pm6$        & 73                  & $3\pm2$                   & 7                            \\
8-9     & $69\pm5$        & 66                  & $1\pm1$                   & 1                            \\
9-10    & $60\pm5$        & 60                  & 0                         & 0                            \\
10-11   & $57\pm5$        & 54                  & 0                         & 0                            \\
11-12   & $50\pm5$        & 50                  & 0                         & 0                            \\
12-13   & $47\pm4$        & 46                  & 0                         & 0                            \\
13-14   & $42\pm4$        & 43                  & 0                         & 0                            \\
14-15   & $40\pm5$        & 40                  & 0                         & 0                            \\
15-16   & $40\pm5$        & 37                  & 0                         & 0                            \\
\bottomrule
\end{tabular}
\label{table:light_cone}
\end{table*}

\section{Conclusions}
\label{sec:conclusions}

We have explored the implication of the Pop III.1 seeding model for cosmological distributions of SMBHs. This is a model that forms all SMBHs with a single mechanism based on the change of protostellar structure in some Pop III stars due to WIMP dark matter particle self annihilation. This leads to reduced ionizing feedback from the protostar and efficient accretion of the baryonic content of the minihalo, thus naturally leading to a characteristic seed mass of $\sim 10^5\:M_\odot$. The model requires the Pop III.1 minihalo to form in relative isolation from other sources. Thus the Pop III.1 seeding model involves all SMBHs forming very early in the Universe, i.e., by $z\sim25$, and with a relatively unclustered initial distribution. Indeed, compared to all other astrophysical models for SMBH formation, the Pop III.1 model involves the earliest and least clustered distribution of seeds. This implies that in the Pop III.1 model, black holes have plenty of time to grow via accretion to explain the known high redshift quasars, without the need of sustained super-Eddington accretion. 

The Pop III.1 model, while being a physical model for the formation of the whole SMBH population, is relatively simple, i.e., with only one free parameter, the isolation distance $d_{\rm iso}$. This means that the model can be easily explored in cosmological volume simulations that resolve minihalos, as was done first in \citetalias{Banik2019}. The constraint of matching an estimate for the local comoving number density of SMBHs, gives quite tight constraints on $d_{\rm iso}\simeq 100\:$kpc (proper distance). This implies most SMBHs formed at $z\simeq 30$, when the isolation distance corresponded to a comoving scale of $\sim 3\:$Mpc. Following on from \citetalias{Banik2019}, we have explored the implications of the Pop III.1 SMBH seeding model down to low redshifts, i.e., all the way to $z=0$, which is important to allow connection to observations, including the HUDF and local galaxy and SMBH populations.  We have also compared this model with another simple seeding scheme,  i.e., the halo mass threshold (HMT) model, that is commonly implemented in cosmological volume simulations.

As presented before, all SMBHs form very early in the universe, and their number density then remains approximately constant after a redshift of $\sim 25$. Only a small fraction of the seeded halos merge with each other by $z=0$. The evolution of the occupation fraction of seeded halos shows a rise to unity for the most massive halos by $z=0$. However, at intermediate redshifts there can be significant fractions of most massive halos that are unseeded.

Our clustering analysis found that Pop III.1 seeded halos show lower levels of clustering compared to random halos with the same mass and number distribution as the seeded halos, at all redshifts. However, to connect this result to observations of AGN \citep[e.g.,][]{Allevato14, He18} requires development of a SMBH growth model, which is planned for a future paper in this series. We also noticed a dip in the clustering of the seeded halos at scales smaller than the isolation distance at the mean formation redshift, which is due to the feedback suppression of the isolation bubbles. This was first discussed at $z=10$ in \citetalias{Banik2019}, and we have shown that this suppression persists even at lower redshift, discernible down to $z\approx 1-2$. 

To compare the clustering of our seeded halos with observational data of galaxies, we turned to the galaxy clustering results from \cite{Zehavi11}. We were able to conclude that the clustering of the seeded halos for 50 and 100 kpc isolation distances are in agreement with the observations, after applying appropriate mass cuts on the halo masses. For comparison with SMBH populations, rather than just galaxies in general, it is clear that having the most complete census in a well-defined, relatively large local volume is highly desirable. Information on this local population is needed both for the total number density and to carry out a clustering analysis of the SMBHs that can be compared to the results of our models. However, obtaining such a census is challenging, given the difficulty of detecting relatively low-mass and faint SMBHs \citep[see, e.g.,][]{Reines16}.
Another promising avenue to be explored relates to the properties of binary AGN and resulting mergers that produce gravitational waves, i.e., sensitive to the extreme end of the clustering signal. These aspects will be considered in detail in forthcoming papers in this series.

Finally, we discussed the potential of using high redshift AGN number counts in the HUDF (or other deep fields) to differentiate among seeding mechanisms and for constraining the value of isolation distance. Detection of just a small number of SMBHs at $z\gtrsim8$ would begin to discriminate between the fiducial HMT scheme and the Pop III.1 model.

\section*{Acknowledgements}

We thank an anonymous referee for helpful comments that improved the paper. We thank Nilanjan Banik for useful discussions. We also thank Yuichi Harikane for helpful discussions and for providing intergrated number density for high redshift AGNs. We also thank Vieri Cammelli and Jacopo Salvalaggio for numerous discussions regarding the simulations and the support of the computing centre of INAF-Osservatorio Astronomico di Trieste, under the coordination of the CHIPP project \cite{Bertocco20, Taffoni20}. JCT acknowledges support from ERC Advanced Grant MSTAR.

\section*{Data Availability}

The data underlying this article will be shared on reasonable request to the corresponding author.

%%%%%%%%%%%%%%%%%%%%%%%%%%%%%%%%%%%%%%%%%%%%%%%%%%



\bibliographystyle{mnras}
\bibliography{bibliography}

\begin{thebibliography}{}
\makeatletter
\relax
\def\mn@urlcharsother{\let\do\@makeother \do\$\do\&\do\#\do\^\do\_\do\%\do\~}
\def\mn@doi{\begingroup\mn@urlcharsother \@ifnextchar [ {\mn@doi@}
  {\mn@doi@[]}}
\def\mn@doi@[#1]#2{\def\@tempa{#1}\ifx\@tempa\@empty \href
  {http://dx.doi.org/#2} {doi:#2}\else \href {http://dx.doi.org/#2} {#1}\fi
  \endgroup}
\def\mn@eprint#1#2{\mn@eprint@#1:#2::\@nil}
\def\mn@eprint@arXiv#1{\href {http://arxiv.org/abs/#1} {{\tt arXiv:#1}}}
\def\mn@eprint@dblp#1{\href {http://dblp.uni-trier.de/rec/bibtex/#1.xml}
  {dblp:#1}}
\def\mn@eprint@#1:#2:#3:#4\@nil{\def\@tempa {#1}\def\@tempb {#2}\def\@tempc
  {#3}\ifx \@tempc \@empty \let \@tempc \@tempb \let \@tempb \@tempa \fi \ifx
  \@tempb \@empty \def\@tempb {arXiv}\fi \@ifundefined
  {mn@eprint@\@tempb}{\@tempb:\@tempc}{\expandafter \expandafter \csname
  mn@eprint@\@tempb\endcsname \expandafter{\@tempc}}}

\bibitem[\protect\citeauthoryear{{Abazajian} et~al.,}{{Abazajian}
  et~al.}{2009}]{Abazajian09}
{Abazajian} K.~N.,  et~al., 2009, \mn@doi [\apjs]
  {10.1088/0067-0049/182/2/543}, \href
  {https://ui.adsabs.harvard.edu/abs/2009ApJS..182..543A} {182, 543}

\bibitem[\protect\citeauthoryear{{Abel}, {Bryan}  \& {Norman}}{{Abel}
  et~al.}{2002}]{Abel02}
{Abel} T.,  {Bryan} G.~L.,   {Norman} M.~L.,  2002, \mn@doi [Science]
  {10.1126/science.295.5552.93}, \href
  {https://ui.adsabs.harvard.edu/abs/2002Sci...295...93A} {295, 93}

\bibitem[\protect\citeauthoryear{{Allevato} et~al.,}{{Allevato}
  et~al.}{2014}]{Allevato14}
{Allevato} V.,  et~al., 2014, \mn@doi [\apj] {10.1088/0004-637X/796/1/4}, \href
  {https://ui.adsabs.harvard.edu/abs/2014ApJ...796....4A} {796, 4}

\bibitem[\protect\citeauthoryear{{Banik}, {Tan}  \& {Monaco}}{{Banik}
  et~al.}{2019}]{Banik2019}
{Banik} N.,  {Tan} J.~C.,   {Monaco} P.,  2019, \mn@doi [\mnras]
  {10.1093/mnras/sty3298}, \href
  {https://ui.adsabs.harvard.edu/abs/2019MNRAS.483.3592B} {483, 3592}

\bibitem[\protect\citeauthoryear{{Barber}, {Schaye}, {Bower}, {Crain},
  {Schaller}  \& {Theuns}}{{Barber} et~al.}{2016}]{Barber16}
{Barber} C.,  {Schaye} J.,  {Bower} R.~G.,  {Crain} R.~A.,  {Schaller} M.,
  {Theuns} T.,  2016, \mn@doi [\mnras] {10.1093/mnras/stw1018}, \href
  {https://ui.adsabs.harvard.edu/abs/2016MNRAS.460.1147B} {460, 1147}

\bibitem[\protect\citeauthoryear{{Beckwith} et~al.,}{{Beckwith}
  et~al.}{2006}]{Beckwith06}
{Beckwith} S. V.~W.,  et~al., 2006, \mn@doi [\aj] {10.1086/507302}, \href
  {https://ui.adsabs.harvard.edu/abs/2006AJ....132.1729B} {132, 1729}

\bibitem[\protect\citeauthoryear{{Begelman}, {Volonteri}  \& {Rees}}{{Begelman}
  et~al.}{2006}]{Begelman06}
{Begelman} M.~C.,  {Volonteri} M.,   {Rees} M.~J.,  2006, \mn@doi [\mnras]
  {10.1111/j.1365-2966.2006.10467.x}, \href
  {https://ui.adsabs.harvard.edu/abs/2006MNRAS.370..289B} {370, 289}

\bibitem[\protect\citeauthoryear{{Bertocco} et~al.,}{{Bertocco}
  et~al.}{2020}]{Bertocco20}
{Bertocco} S.,  et~al., 2020, in {Pizzo} R.,  {Deul} E.~R.,  {Mol} J.~D.,  {de
  Plaa} J.,   {Verkouter} H.,  eds,  Astronomical Society of the Pacific
  Conference Series Vol. 527, Astronomical Data Analysis Software and Systems
  XXIX. p.~303 (\mn@eprint {arXiv} {1912.05340}),
  \mn@doi{10.48550/arXiv.1912.05340}

\bibitem[\protect\citeauthoryear{{Bhowmick}, {Blecha}, {Torrey}, {Kelley},
  {Vogelsberger}, {Nelson}, {Weinberger}  \& {Hernquist}}{{Bhowmick}
  et~al.}{2022a}]{Bhowmick22a}
{Bhowmick} A.~K.,  {Blecha} L.,  {Torrey} P.,  {Kelley} L.~Z.,  {Vogelsberger}
  M.,  {Nelson} D.,  {Weinberger} R.,   {Hernquist} L.,  2022a, \mn@doi
  [\mnras] {10.1093/mnras/stab3439}, \href
  {https://ui.adsabs.harvard.edu/abs/2022MNRAS.510..177B} {510, 177}

\bibitem[\protect\citeauthoryear{{Bhowmick} et~al.,}{{Bhowmick}
  et~al.}{2022b}]{Bhowmick22b}
{Bhowmick} A.~K.,  et~al., 2022b, \mn@doi [\mnras] {10.1093/mnras/stac2238},
  \href {https://ui.adsabs.harvard.edu/abs/2022MNRAS.516..138B} {516, 138}

\bibitem[\protect\citeauthoryear{{Boekholt}, {Schleicher}, {Fellhauer},
  {Klessen}, {Reinoso}, {Stutz}  \& {Haemmerl{\'e}}}{{Boekholt}
  et~al.}{2018}]{Boekholt18}
{Boekholt} T.~C.~N.,  {Schleicher} D.~R.~G.,  {Fellhauer} M.,  {Klessen} R.~S.,
   {Reinoso} B.,  {Stutz} A.~M.,   {Haemmerl{\'e}} L.,  2018, \mn@doi [\mnras]
  {10.1093/mnras/sty208}, \href
  {https://ui.adsabs.harvard.edu/abs/2018MNRAS.476..366B} {476, 366}

\bibitem[\protect\citeauthoryear{{Bromm} \& {Loeb}}{{Bromm} \&
  {Loeb}}{2003}]{Bromm03}
{Bromm} V.,  {Loeb} A.,  2003, \mn@doi [\apj] {10.1086/377529}, \href
  {https://ui.adsabs.harvard.edu/abs/2003ApJ...596...34B} {596, 34}

\bibitem[\protect\citeauthoryear{{Bromm}, {Coppi}  \& {Larson}}{{Bromm}
  et~al.}{2002}]{Bromm02}
{Bromm} V.,  {Coppi} P.~S.,   {Larson} R.~B.,  2002, \mn@doi [\apj]
  {10.1086/323947}, \href
  {https://ui.adsabs.harvard.edu/abs/2002ApJ...564...23B} {564, 23}

\bibitem[\protect\citeauthoryear{{Chon} \& {Omukai}}{{Chon} \&
  {Omukai}}{2020}]{Chon20}
{Chon} S.,  {Omukai} K.,  2020, \mn@doi [\mnras] {10.1093/mnras/staa863}, \href
  {https://ui.adsabs.harvard.edu/abs/2020MNRAS.494.2851C} {494, 2851}

\bibitem[\protect\citeauthoryear{{Chon}, {Hirano}, {Hosokawa}  \&
  {Yoshida}}{{Chon} et~al.}{2016}]{Chon16}
{Chon} S.,  {Hirano} S.,  {Hosokawa} T.,   {Yoshida} N.,  2016, \mn@doi [\apj]
  {10.3847/0004-637X/832/2/134}, \href
  {https://ui.adsabs.harvard.edu/abs/2016ApJ...832..134C} {832, 134}

\bibitem[\protect\citeauthoryear{{Comparat}, {Prada}, {Yepes}  \&
  {Klypin}}{{Comparat} et~al.}{2017}]{Comparat17}
{Comparat} J.,  {Prada} F.,  {Yepes} G.,   {Klypin} A.,  2017, \mn@doi [\mnras]
  {10.1093/mnras/stx1183}, \href
  {https://ui.adsabs.harvard.edu/abs/2017MNRAS.469.4157C} {469, 4157}

\bibitem[\protect\citeauthoryear{{Crocce}, {Fosalba}, {Castander}  \&
  {Gazta{\~n}aga}}{{Crocce} et~al.}{2010}]{Crocce10}
{Crocce} M.,  {Fosalba} P.,  {Castander} F.~J.,   {Gazta{\~n}aga} E.,  2010,
  \mn@doi [\mnras] {10.1111/j.1365-2966.2009.16194.x}, \href
  {https://ui.adsabs.harvard.edu/abs/2010MNRAS.403.1353C} {403, 1353}

\bibitem[\protect\citeauthoryear{{Di Matteo}, {Colberg}, {Springel},
  {Hernquist}  \& {Sijacki}}{{Di Matteo} et~al.}{2008}]{DiMatteo08}
{Di Matteo} T.,  {Colberg} J.,  {Springel} V.,  {Hernquist} L.,   {Sijacki} D.,
   2008, \mn@doi [\apj] {10.1086/524921}, \href
  {https://ui.adsabs.harvard.edu/abs/2008ApJ...676...33D} {676, 33}

\bibitem[\protect\citeauthoryear{{Diemer}}{{Diemer}}{2018}]{Diemer18}
{Diemer} B.,  2018, \mn@doi [\apjs] {10.3847/1538-4365/aaee8c}, \href
  {https://ui.adsabs.harvard.edu/abs/2018ApJS..239...35D} {239, 35}

\bibitem[\protect\citeauthoryear{{Ebisuzaki}}{{Ebisuzaki}}{2003}]{Ebisuzaki03}
{Ebisuzaki} T.,  2003, in {Makino} J.,  {Hut} P.,  eds,  IAU Symposium Vol.
  208, Astrophysical Supercomputing using Particle Simulations. p.~157

\bibitem[\protect\citeauthoryear{{Ellis} et~al.,}{{Ellis}
  et~al.}{2013}]{Ellis13}
{Ellis} R.~S.,  et~al., 2013, \mn@doi [\apjl] {10.1088/2041-8205/763/1/L7},
  \href {https://ui.adsabs.harvard.edu/abs/2013ApJ...763L...7E} {763, L7}

\bibitem[\protect\citeauthoryear{{Feng}, {Yu}  \& {Zhong}}{{Feng}
  et~al.}{2021}]{Feng21}
{Feng} W.-X.,  {Yu} H.-B.,   {Zhong} Y.-M.,  2021, \mn@doi [\apjl]
  {10.3847/2041-8213/ac04b0}, \href
  {https://ui.adsabs.harvard.edu/abs/2021ApJ...914L..26F} {914, L26}

\bibitem[\protect\citeauthoryear{Freese, Ilie, Spolyar, Valluri  \&
  Bodenheimer}{Freese et~al.}{2010}]{F10}
Freese K.,  Ilie C.,  Spolyar D.,  Valluri M.,   Bodenheimer P.,  2010, \mn@doi
  [\apj] {10.1088/0004-637x/716/2/1397}, 716, 1397

\bibitem[\protect\citeauthoryear{{Graham}}{{Graham}}{2016}]{Graham16}
{Graham} A.~W.,  2016, in {Laurikainen} E.,  {Peletier} R.,   {Gadotti} D.,
  eds,  Astrophysics and Space Science Library Vol. 418, Galactic Bulges.
  p.~263 (\mn@eprint {arXiv} {1501.02937}),
  \mn@doi{10.1007/978-3-319-19378-6_11}

\bibitem[\protect\citeauthoryear{{Greif} \& {Bromm}}{{Greif} \&
  {Bromm}}{2006}]{Greif2006}
{Greif} T.~H.,  {Bromm} V.,  2006, \mn@doi [\mnras]
  {10.1111/j.1365-2966.2006.11017.x}, \href
  {https://ui.adsabs.harvard.edu/abs/2006MNRAS.373..128G} {373, 128}

\bibitem[\protect\citeauthoryear{{Harikane} et~al.,}{{Harikane}
  et~al.}{2023}]{Harikane23}
{Harikane} Y.,  et~al., 2023, \mn@doi [arXiv e-prints]
  {10.48550/arXiv.2303.11946}, \href
  {https://ui.adsabs.harvard.edu/abs/2023arXiv230311946H} {p. arXiv:2303.11946}

\bibitem[\protect\citeauthoryear{{He} et~al.,}{{He} et~al.}{2018}]{He18}
{He} W.,  et~al., 2018, \mn@doi [\pasj] {10.1093/pasj/psx129}, \href
  {https://ui.adsabs.harvard.edu/abs/2018PASJ...70S..33H} {70, S33}

\bibitem[\protect\citeauthoryear{{Hosokawa}, {Omukai}, {Yoshida}  \&
  {Yorke}}{{Hosokawa} et~al.}{2011}]{Hosokawa11}
{Hosokawa} T.,  {Omukai} K.,  {Yoshida} N.,   {Yorke} H.~W.,  2011, \mn@doi
  [Science] {10.1126/science.1207433}, \href
  {https://ui.adsabs.harvard.edu/abs/2011Sci...334.1250H} {334, 1250}

\bibitem[\protect\citeauthoryear{{Jeon}, {Liu}, {Bromm}  \&
  {Finkelstein}}{{Jeon} et~al.}{2023}]{Jeon23}
{Jeon} J.,  {Liu} B.,  {Bromm} V.,   {Finkelstein} S.~L.,  2023, \mn@doi [arXiv
  e-prints] {10.48550/arXiv.2304.07369}, \href
  {https://ui.adsabs.harvard.edu/abs/2023arXiv230407369J} {p. arXiv:2304.07369}

\bibitem[\protect\citeauthoryear{{Kohri}, {Sekiguchi}  \& {Wang}}{{Kohri}
  et~al.}{2022}]{Kohri22}
{Kohri} K.,  {Sekiguchi} T.,   {Wang} S.,  2022, \mn@doi [\prd]
  {10.1103/PhysRevD.106.043539}, \href
  {https://ui.adsabs.harvard.edu/abs/2022PhRvD.106d3539K} {106, 043539}

\bibitem[\protect\citeauthoryear{{Kroupa}, {Subr}, {Jerabkova}  \&
  {Wang}}{{Kroupa} et~al.}{2020}]{Kroupa20}
{Kroupa} P.,  {Subr} L.,  {Jerabkova} T.,   {Wang} L.,  2020, \mn@doi [\mnras]
  {10.1093/mnras/staa2276}, \href
  {https://ui.adsabs.harvard.edu/abs/2020MNRAS.498.5652K} {498, 5652}

\bibitem[\protect\citeauthoryear{{Lodato} \& {Natarajan}}{{Lodato} \&
  {Natarajan}}{2006}]{Lodato06}
{Lodato} G.,  {Natarajan} P.,  2006, \mn@doi [\mnras]
  {10.1111/j.1365-2966.2006.10801.x}, \href
  {https://ui.adsabs.harvard.edu/abs/2006MNRAS.371.1813L} {371, 1813}

\bibitem[\protect\citeauthoryear{{Lusso}, {Valiante}  \& {Vito}}{{Lusso}
  et~al.}{2022}]{Lusso22}
{Lusso} E.,  {Valiante} R.,   {Vito} F.,  2022, \mn@doi [arXiv e-prints]
  {10.48550/arXiv.2205.15349}, \href
  {https://ui.adsabs.harvard.edu/abs/2022arXiv220515349L} {p. arXiv:2205.15349}

\bibitem[\protect\citeauthoryear{{Maio}, {Borgani}, {Ciardi}  \&
  {Petkova}}{{Maio} et~al.}{2019}]{Maio19}
{Maio} U.,  {Borgani} S.,  {Ciardi} B.,   {Petkova} M.,  2019, \mn@doi [\pasa]
  {10.1017/pasa.2019.10}, \href
  {https://ui.adsabs.harvard.edu/abs/2019PASA...36...20M} {36, e020}

\bibitem[\protect\citeauthoryear{McKee \& Tan}{McKee \& Tan}{2008}]{MT08}
McKee C.~F.,  Tan J.~C.,  2008, \mn@doi [\apj] {10.1086/587434}, 681, 771

\bibitem[\protect\citeauthoryear{Monaco, Theuns  \& Taffoni}{Monaco
  et~al.}{2002}]{MTT02}
Monaco P.,  Theuns T.,   Taffoni G.,  2002, \mn@doi [\mnras]
  {10.1046/j.1365-8711.2002.05162.x}, 331, 587

\bibitem[\protect\citeauthoryear{{Montero}, {Janka}  \& {M{\"u}ller}}{{Montero}
  et~al.}{2012}]{Montero12}
{Montero} P.~J.,  {Janka} H.-T.,   {M{\"u}ller} E.,  2012, \mn@doi [\apj]
  {10.1088/0004-637X/749/1/37}, \href
  {https://ui.adsabs.harvard.edu/abs/2012ApJ...749...37M} {749, 37}

\bibitem[\protect\citeauthoryear{{Moutarde}, {Alimi}, {Bouchet}, {Pellat}  \&
  {Ramani}}{{Moutarde} et~al.}{1991}]{M91}
{Moutarde} F.,  {Alimi} J.~M.,  {Bouchet} F.~R.,  {Pellat} R.,   {Ramani} A.,
  1991, \mn@doi [\apj] {10.1086/170728}, \href
  {https://ui.adsabs.harvard.edu/abs/1991ApJ...382..377M} {382, 377}

\bibitem[\protect\citeauthoryear{{Munari}, {Monaco}, {Sefusatti}, {Castorina},
  {Mohammad}, {Anselmi}  \& {Borgani}}{{Munari} et~al.}{2017}]{Munari17}
{Munari} E.,  {Monaco} P.,  {Sefusatti} E.,  {Castorina} E.,  {Mohammad} F.~G.,
   {Anselmi} S.,   {Borgani} S.,  2017, \mn@doi [\mnras]
  {10.1093/mnras/stw3085}, \href
  {https://ui.adsabs.harvard.edu/abs/2017MNRAS.465.4658M} {465, 4658}

\bibitem[\protect\citeauthoryear{{Nakajima}, {Ouchi}, {Isobe}, {Harikane},
  {Zhang}, {Ono}, {Umeda}  \& {Oguri}}{{Nakajima} et~al.}{2023}]{Nakajima23}
{Nakajima} K.,  {Ouchi} M.,  {Isobe} Y.,  {Harikane} Y.,  {Zhang} Y.,  {Ono}
  Y.,  {Umeda} H.,   {Oguri} M.,  2023, \mn@doi [arXiv e-prints]
  {10.48550/arXiv.2301.12825}, \href
  {https://ui.adsabs.harvard.edu/abs/2023arXiv230112825N} {p. arXiv:2301.12825}

\bibitem[\protect\citeauthoryear{Natarajan, Tan  \&
  O{\textquotesingle}Shea}{Natarajan et~al.}{2009}]{NTO09}
Natarajan A.,  Tan J.~C.,   O{\textquotesingle}Shea B.~W.,  2009, \mn@doi
  [\apj] {10.1088/0004-637x/692/1/574}, 692, 574

\bibitem[\protect\citeauthoryear{Norberg et~al.,}{Norberg et~al.}{2002}]{N02}
Norberg P.,  et~al., 2002, \mn@doi [\mnras] {10.1046/j.1365-8711.2002.05831.x},
  336, 907

\bibitem[\protect\citeauthoryear{{O'Shea}, {Abel}, {Whalen}  \&
  {Norman}}{{O'Shea} et~al.}{2005}]{Oshea05}
{O'Shea} B.~W.,  {Abel} T.,  {Whalen} D.,   {Norman} M.~L.,  2005, \mn@doi
  [\apjl] {10.1086/432683}, \href
  {https://ui.adsabs.harvard.edu/abs/2005ApJ...628L...5O} {628, L5}

\bibitem[\protect\citeauthoryear{{Planck Collaboration} et~al.,}{{Planck
  Collaboration} et~al.}{2020}]{Planck20}
{Planck Collaboration} et~al., 2020, \mn@doi [\aap]
  {10.1051/0004-6361/201833910}, \href
  {https://ui.adsabs.harvard.edu/abs/2020A&A...641A...6P} {641, A6}

\bibitem[\protect\citeauthoryear{{Portegies Zwart}, {Baumgardt}, {Hut},
  {Makino}  \& {McMillan}}{{Portegies Zwart} et~al.}{2004}]{Zwart04}
{Portegies Zwart} S.~F.,  {Baumgardt} H.,  {Hut} P.,  {Makino} J.,   {McMillan}
  S. L.~W.,  2004, \mn@doi [\nat] {10.1038/nature02448}, \href
  {https://ui.adsabs.harvard.edu/abs/2004Natur.428..724P} {428, 724}

\bibitem[\protect\citeauthoryear{{Rees}}{{Rees}}{1978}]{Rees78}
{Rees} M.~J.,  1978, The Observatory, \href
  {https://ui.adsabs.harvard.edu/abs/1978Obs....98..210R} {98, 210}

\bibitem[\protect\citeauthoryear{{Regan}, {Wise}, {Woods}, {Downes}, {O'Shea}
  \& {Norman}}{{Regan} et~al.}{2020}]{Regan20}
{Regan} J.~A.,  {Wise} J.~H.,  {Woods} T.~E.,  {Downes} T.~P.,  {O'Shea} B.~W.,
    {Norman} M.~L.,  2020, \mn@doi [The Open Journal of Astrophysics]
  {10.21105/astro.2008.08090}, \href
  {https://ui.adsabs.harvard.edu/abs/2020OJAp....3E..15R} {3, 15}

\bibitem[\protect\citeauthoryear{{Reines} \& {Comastri}}{{Reines} \&
  {Comastri}}{2016}]{Reines16}
{Reines} A.~E.,  {Comastri} A.,  2016, \mn@doi [\pasa] {10.1017/pasa.2016.46},
  \href {https://ui.adsabs.harvard.edu/abs/2016PASA...33...54R} {33, e054}

\bibitem[\protect\citeauthoryear{Rindler-Daller, Montgomery, Freese, Winget  \&
  Paxton}{Rindler-Daller et~al.}{2015}]{RD15}
Rindler-Daller T.,  Montgomery M.~H.,  Freese K.,  Winget D.~E.,   Paxton B.,
  2015, \mn@doi [\apj] {10.1088/0004-637x/799/2/210}, 799, 210

\bibitem[\protect\citeauthoryear{{Shang}, {Bryan}  \& {Haiman}}{{Shang}
  et~al.}{2010}]{Shang10}
{Shang} C.,  {Bryan} G.~L.,   {Haiman} Z.,  2010, \mn@doi [\mnras]
  {10.1111/j.1365-2966.2009.15960.x}, \href
  {https://ui.adsabs.harvard.edu/abs/2010MNRAS.402.1249S} {402, 1249}

\bibitem[\protect\citeauthoryear{{Shinohara}, {He}, {Matsuoka}, {Nagao},
  {Suyama}  \& {Takahashi}}{{Shinohara} et~al.}{2023}]{Shinohara23}
{Shinohara} T.,  {He} W.,  {Matsuoka} Y.,  {Nagao} T.,  {Suyama} T.,
  {Takahashi} T.,  2023, \mn@doi [arXiv e-prints] {10.48550/arXiv.2304.08153},
  \href {https://ui.adsabs.harvard.edu/abs/2023arXiv230408153S} {p.
  arXiv:2304.08153}

\bibitem[\protect\citeauthoryear{{Sijacki}, {Springel}, {Di Matteo}  \&
  {Hernquist}}{{Sijacki} et~al.}{2007}]{Sijacki07}
{Sijacki} D.,  {Springel} V.,  {Di Matteo} T.,   {Hernquist} L.,  2007, \mn@doi
  [\mnras] {10.1111/j.1365-2966.2007.12153.x}, \href
  {https://ui.adsabs.harvard.edu/abs/2007MNRAS.380..877S} {380, 877}

\bibitem[\protect\citeauthoryear{Sijacki, Vogelsberger, Genel, Springel,
  Torrey, Snyder, Nelson  \& Hernquist}{Sijacki et~al.}{2015}]{Sijaki15}
Sijacki D.,  Vogelsberger M.,  Genel S.,  Springel V.,  Torrey P.,  Snyder
  G.~F.,  Nelson D.,   Hernquist L.,  2015, \mn@doi [\mnras]
  {10.1093/mnras/stv1340}, 452, 575

\bibitem[\protect\citeauthoryear{{Sinha} \& {Garrison}}{{Sinha} \&
  {Garrison}}{2020}]{Sinha20}
{Sinha} M.,  {Garrison} L.~H.,  2020, \mn@doi [\mnras] {10.1093/mnras/stz3157},
  \href {https://ui.adsabs.harvard.edu/abs/2020MNRAS.491.3022S} {491, 3022}

\bibitem[\protect\citeauthoryear{Spolyar, Freese  \& Gondolo}{Spolyar
  et~al.}{2008}]{SFG08}
Spolyar D.,  Freese K.,   Gondolo P.,  2008, \mn@doi [Phys. Rev. Lett.]
  {10.1103/PhysRevLett.100.051101}, 100, 051101

\bibitem[\protect\citeauthoryear{{Susa}, {Hasegawa}  \& {Tominaga}}{{Susa}
  et~al.}{2014}]{Susa14}
{Susa} H.,  {Hasegawa} K.,   {Tominaga} N.,  2014, \mn@doi [\apj]
  {10.1088/0004-637X/792/1/32}, \href
  {https://ui.adsabs.harvard.edu/abs/2014ApJ...792...32S} {792, 32}

\bibitem[\protect\citeauthoryear{{Taffoni}, {Becciani}, {Garilli}, {Maggio},
  {Pasian}, {Umana}, {Smareglia}  \& {Vitello}}{{Taffoni}
  et~al.}{2020}]{Taffoni20}
{Taffoni} G.,  {Becciani} U.,  {Garilli} B.,  {Maggio} G.,  {Pasian} F.,
  {Umana} G.,  {Smareglia} R.,   {Vitello} F.,  2020, in {Pizzo} R.,  {Deul}
  E.~R.,  {Mol} J.~D.,  {de Plaa} J.,   {Verkouter} H.,  eds,  Astronomical
  Society of the Pacific Conference Series Vol. 527, Astronomical Data Analysis
  Software and Systems XXIX. p.~307 (\mn@eprint {arXiv} {2002.01283}),
  \mn@doi{10.48550/arXiv.2002.01283}

\bibitem[\protect\citeauthoryear{{Tagawa}, {Haiman}  \& {Kocsis}}{{Tagawa}
  et~al.}{2020}]{Tagawa20}
{Tagawa} H.,  {Haiman} Z.,   {Kocsis} B.,  2020, \mn@doi [\apj]
  {10.3847/1538-4357/ab7922}, \href
  {https://ui.adsabs.harvard.edu/abs/2020ApJ...892...36T} {892, 36}

\bibitem[\protect\citeauthoryear{{Tan} \& {McKee}}{{Tan} \&
  {McKee}}{2004}]{Tan04}
{Tan} J.~C.,  {McKee} C.~F.,  2004, \mn@doi [\apj] {10.1086/381490}, \href
  {https://ui.adsabs.harvard.edu/abs/2004ApJ...603..383T} {603, 383}

\bibitem[\protect\citeauthoryear{{Trebitsch} et~al.,}{{Trebitsch}
  et~al.}{2021}]{Trebitsch21}
{Trebitsch} M.,  et~al., 2021, \mn@doi [\aap] {10.1051/0004-6361/202037698},
  \href {https://ui.adsabs.harvard.edu/abs/2021A&A...653A.154T} {653, A154}

\bibitem[\protect\citeauthoryear{{Tremmel}, {Karcher}, {Governato},
  {Volonteri}, {Quinn}, {Pontzen}, {Anderson}  \& {Bellovary}}{{Tremmel}
  et~al.}{2017}]{Tremmel17}
{Tremmel} M.,  {Karcher} M.,  {Governato} F.,  {Volonteri} M.,  {Quinn} T.~R.,
  {Pontzen} A.,  {Anderson} L.,   {Bellovary} J.,  2017, \mn@doi [\mnras]
  {10.1093/mnras/stx1160}, \href
  {https://ui.adsabs.harvard.edu/abs/2017MNRAS.470.1121T} {470, 1121}

\bibitem[\protect\citeauthoryear{{Vogelsberger} et~al.,}{{Vogelsberger}
  et~al.}{2014}]{Vogelsberger14}
{Vogelsberger} M.,  et~al., 2014, \mn@doi [\mnras] {10.1093/mnras/stu1536},
  \href {https://ui.adsabs.harvard.edu/abs/2014MNRAS.444.1518V} {444, 1518}

\bibitem[\protect\citeauthoryear{{Volonteri}, {Dubois}, {Pichon}  \&
  {Devriendt}}{{Volonteri} et~al.}{2016}]{Volonteri16}
{Volonteri} M.,  {Dubois} Y.,  {Pichon} C.,   {Devriendt} J.,  2016, \mn@doi
  [\mnras] {10.1093/mnras/stw1123}, \href
  {https://ui.adsabs.harvard.edu/abs/2016MNRAS.460.2979V} {460, 2979}

\bibitem[\protect\citeauthoryear{{Volonteri}, {Habouzit}  \&
  {Colpi}}{{Volonteri} et~al.}{2021}]{Volonteri21}
{Volonteri} M.,  {Habouzit} M.,   {Colpi} M.,  2021, \mn@doi [Nature Reviews
  Physics] {10.1038/s42254-021-00364-9}, \href
  {https://ui.adsabs.harvard.edu/abs/2021NatRP...3..732V} {3, 732}

\bibitem[\protect\citeauthoryear{{Wang} et~al.,}{{Wang} et~al.}{2021}]{Wang21}
{Wang} F.,  et~al., 2021, \mn@doi [\apjl] {10.3847/2041-8213/abd8c6}, \href
  {https://ui.adsabs.harvard.edu/abs/2021ApJ...907L...1W} {907, L1}

\bibitem[\protect\citeauthoryear{{Wise}, {Regan}, {O'Shea}, {Norman}, {Downes}
  \& {Xu}}{{Wise} et~al.}{2019}]{Wise2019}
{Wise} J.~H.,  {Regan} J.~A.,  {O'Shea} B.~W.,  {Norman} M.~L.,  {Downes}
  T.~P.,   {Xu} H.,  2019, \mn@doi [\nat] {10.1038/s41586-019-0873-4}, \href
  {https://ui.adsabs.harvard.edu/abs/2019Natur.566...85W} {566, 85}

\bibitem[\protect\citeauthoryear{{Yang} et~al.,}{{Yang} et~al.}{2020}]{Yang20}
{Yang} J.,  et~al., 2020, \mn@doi [\apjl] {10.3847/2041-8213/ab9c26}, \href
  {https://ui.adsabs.harvard.edu/abs/2020ApJ...897L..14Y} {897, L14}

\bibitem[\protect\citeauthoryear{{York} et~al.,}{{York} et~al.}{2000}]{York00}
{York} D.~G.,  et~al., 2000, \mn@doi [\aj] {10.1086/301513}, \href
  {https://ui.adsabs.harvard.edu/abs/2000AJ....120.1579Y} {120, 1579}

\bibitem[\protect\citeauthoryear{{Zehavi} et~al.,}{{Zehavi}
  et~al.}{2011}]{Zehavi11}
{Zehavi} I.,  et~al., 2011, \mn@doi [\apj] {10.1088/0004-637X/736/1/59}, \href
  {https://ui.adsabs.harvard.edu/abs/2011ApJ...736...59Z} {736, 59}

\makeatother
\end{thebibliography}

%\begin{thebibliography}{99}
%\end{thebibliography}

%%%%%%%%%%%%%%%%%%%%%%%%%%%%%%%%%%%%%%%%%%%%%%%%%%

%%%%%%%%%%%%%%%%% APPENDICES %%%%%%%%%%%%%%%%%%%%%

\appendix

% If you want to present additional material which would interrupt the flow of the main paper,
% it can be placed in an Appendix which appears after the list of references.

\section{Matching full- and low-resolution {\sc pinocchio} runs}

\begin{figure}
    \centering
    \includegraphics[width=0.46\textwidth]{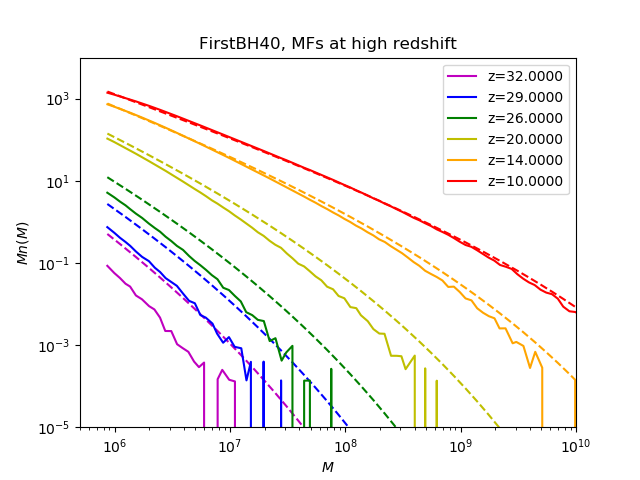}
    \caption{Halo mass function of the full-resolution box at high redshift. Lines are color-coded in redshift (see legend). Solid lines refer to {\sc pinocchio} catalogs; dashed lines to the \citet{Crocce10} analytic fit.}
    \label{fig:mf_highz}
\end{figure}

We run the 59.7 Mpc box at the full resolution of $4096^3$ particles and at a lower resolution of $1024^3$ particles. These resolutions correspond to particle masses of $1.23 \times 10^5 M_\odot$ and $7.87\times 10^6 M_\odot$. The minimum mass for halos has been set to 10 particles in both cases. Figure \ref{fig:mf_highz} shows the mass function of the full-resolution box at high redshift, where it is evident that the early growth of massive halos is slower than in a universal model (in this case the fit to the friends-of-friends halo mass function of \cite{Crocce10}). We stress that there is no reason to believe that this analytic fit is accurate at such low masses, but we conservatively assume that the disagreement is due to an inaccuracy of {\sc pinocchio}.

\begin{figure}
    \centering
    \includegraphics[width=0.46\textwidth]{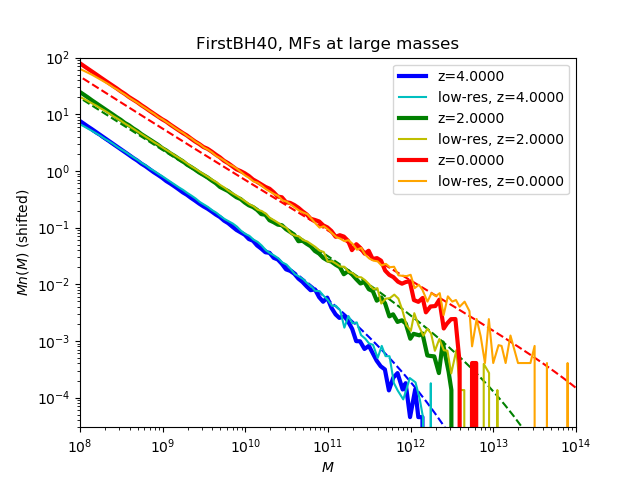}
    \caption{Halo mass function of the full- (thick solid lines) and low-resolution (thin solid lines) boxes at low redshift. Lines are color-coded in redshift (see legend). Dashed lines are the \citet{Crocce10} analytic fit.}
    \label{fig:mf_highmass}
\end{figure}

\begin{figure}
    \centering
    \includegraphics[width=0.46\textwidth]{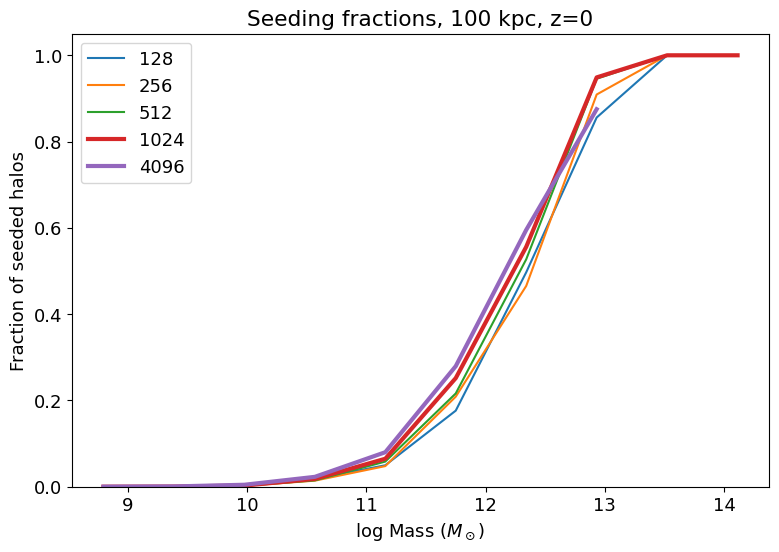}
    \caption{Fraction of halos of a given mass that contain a seed SMBH, for $d_{\rm iso}=100$ kpc. Resolution is color-coded (see legend). Thicker lines emphasize the full-resolution (4096) and low-resolution (1024) runs.}
    \label{fig:seeding}
\end{figure}

Figure \ref{fig:mf_highmass} shows the halo mass function for the low-resolution box (thin lines) and the full-resolution run (thick lines). At high masses the agreement of the high-resolution box with the analytic prediction is poor, while this is not the case for the low-resolution run where the box has not been divided into different domains. Figure \ref{fig:seeding} shows the consistency of the seeding fraction among the high-resolution box and a set of lower and lower resolution runs, where seeding of halos is decided by checking which particle in Lagrangian space contains the halos that is seeded in the full resolution box.

\section{Large-scale clustering modes}
\label{appendix:ls_corr}

When we use the estimators such as the \textsc{corrfunc} library to find the auto correlation of halos in our 59.7 Mpc box, the correlation function only contains the clustering modes smaller than the box size. If we want to make a simplistic comparison of our results with a large survey which sampled a much larger volume, we can do so by analytically adding the larger scale clustering modes. To understand how we achieve this, we examine the analytic expression for calculating the 3D 2pcf for halos for the entire volume of the Universe:
\begin{equation}
    \xi_{hh}(r)=\frac{1}{2\pi^2}\int_0^\infty dk k^2 b_h^2P(k)\frac{\sin{kr}}{kr},
\end{equation}
where $\xi_{hh}$ is the correlation function of halos, $b_h$ is the halo bias, and $P(k)$ is the matter power spectrum. This integral can be split in two parts:
\begin{equation}
\begin{aligned}
    \xi_{hh}(r) = & \underbrace{\frac{1}{2\pi^2}\int_0^{k_{\rm box}} dk k^2 b_h^2P(k)\frac{\sin{kr}}{kr}}_{\text{Large scale contribution } \xi_{\rm LS}(r)} \\
    & + \underbrace{\frac{1}{2\pi^2}\int^\infty_{k_{\rm box}} dk k^2 b_h^2P(k)\frac{\sin{kr}}{kr}}_{\text{\textsc{pinocchio} contribution }\xi_{\rm PIN}(r)} \\
    & = \xi_{\rm LS}(r) + \xi_{\rm PIN}(r),
\end{aligned}
\end{equation}
where $k_{\rm box}=2\pi/L_{\rm box}$, with $L_{\rm box}=59.7$ Mpc in our box. The large scale contribution refers to the clustering modes of radial scale going from $L_{\rm box}$ to infinity, and the \textsc{pinocchio} contribution refers to all the modes of radial scale from 0 to $L_{\rm box}$. Since the correlation estimator returns $\xi_{\rm PIN}$, we calculated the large scale contribution by using the linear matter power spectrum from \textsc{camb} python library and halo bias from \textsc{colossus} python library \citep{Diemer18}, using the bias model of \cite{Comparat17}, and then numerically integrated the power spectrum to obtain $\xi_{\rm LS}$.

\begin{figure*}
    \centering
    \includegraphics[scale=0.59]{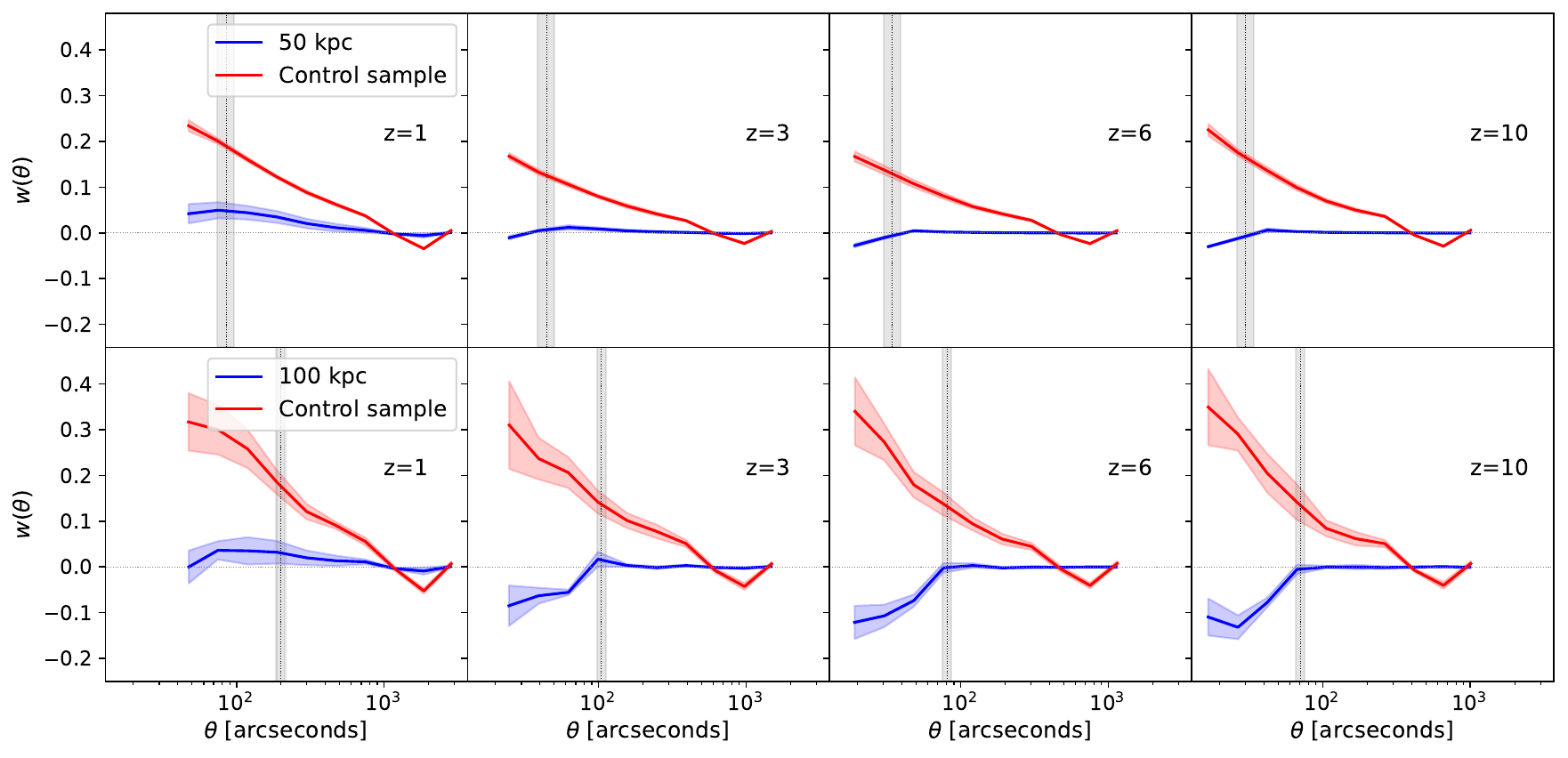}
    \caption{Evolution of angular correlation function for 50 and 100 kpc isolation distances. The large scale modes are not added in the evaluation of this function. The labels are the same as in Figure \ref{fig:clus_evolution}.}
    \label{fig:ang_evol}
\end{figure*}

To make a direct continuation of the angular clustering as shown in \cite{Banik2019} (their Figure 10), we present the angular clustering evolution of seeded halos in Figure \ref{fig:ang_evol} without the large scale corrections added. This figure and  Figure \ref{fig:clus_evolution} essentially show the same information, with the only difference that the figure presented here is in angular scale, and without the large scale modes.

%%%%%%%%%%%%%%%%%%%%%%%%%%%%%%%%%%%%%%%%%%%%%%%%%%

% Don't change these lines
\bsp	% typesetting comment
\label{lastpage}
\end{document}